\documentclass[twocolumn]{aastex631}

\shorttitle{Optical-NIR dark strongly lensed galaxy}
\shortauthors{Giulietti et al.}

\graphicspath{{./}{figures/}}
\usepackage{tabularx}
\usepackage{amsmath}

\begin{document}

\title{ALMA resolves the first strongly-lensed Optical/NIR-dark galaxy}


\author[0000-0002-1847-4496]{Marika Giulietti}
\affiliation{SISSA, Via Bonomea 265, I-34136 Trieste, Italy}
\affiliation{INAF - Osservatorio di Astrofisica e Scienza dello Spazio, Via Gobetti 93/3, I-40129, Bologna, Italy}

\author[0000-0002-4882-1735]{Andrea Lapi}
\affiliation{SISSA, Via Bonomea 265, I-34136 Trieste, Italy}\affiliation{IFPU - Institute for fundamental physics of the Universe, Via Beirut 2, 34014 Trieste, Italy}\affiliation{INFN-Sezione di Trieste, via Valerio 2, 34127 Trieste,  Italy}\affiliation{INAF/IRA, Istituto di Radioastronomia, Via Piero Gobetti 101, 40129 Bologna, Italy}

\author[0000-0002-0375-8330]{Marcella Massardi}
\affiliation{INAF/IRA, Istituto di Radioastronomia, Via Piero Gobetti 101, 40129 Bologna, Italy}\affiliation{INAF, Istituto di Radioastronomia - Italian ARC, Via Piero Gobetti 101, I-40129 Bologna, Italy}\affiliation{SISSA, Via Bonomea 265, I-34136 Trieste, Italy}

\author[0000-0002-6444-8547]{Meriem Behiri}
\affiliation{SISSA, Via Bonomea 265, I-34136 Trieste, Italy}

\author[0000-0001-9301-5209]{Martina Torsello}
\affiliation{SISSA, Via Bonomea 265, I-34136 Trieste, Italy}

\author[0000-0002-9948-0897]{Quirino D'Amato}
\affiliation{SISSA, Via Bonomea 265, I-34136 Trieste, Italy}
\affiliation{INAF - Osservatorio di Astrofisica e Scienza dello Spazio, Via Gobetti 93/3, I-40129, Bologna, Italy}

\author[0000-0002-3515-6801]{Tommaso Ronconi}
\affiliation{SISSA, Via Bonomea 265, I-34136 Trieste, Italy}
\affiliation{IFPU, Via Beirut 2, I-34014 Trieste, Italy}
\affiliation{INFN-Sezione di Trieste, via Valerio 2, I-34127 Trieste, Italy}

\author[0000-0001-9808-0843]{Francesca Perrotta}
\affiliation{SISSA, Via Bonomea 265, I-34136 Trieste, Italy}

\author[0000-0002-7922-8440]{Alessandro Bressan}
\affiliation{SISSA, Via Bonomea 265, I-34136 Trieste, Italy}



\begin{abstract}
We present high-resolution ($\lesssim0.1$arcsec) ALMA observations of the strongly-lensed galaxy HATLASJ113526.2-01460 at redshift $z\sim3.1$ discovered in the Gama 12$^{\rm th}$ field of the \textit{Herschel}-ATLAS survey. The gravitationally lensed system is remarkably peculiar in that neither the background source nor the foreground lens show a clearly detected optical/NIR emission. We perform accurate lens modeling and source morphology reconstruction in three different (sub-)mm continuum bands, and in the C[II] and CO(8-7) spectral lines. The modeling indicates a foreground lensing (likely elliptical) galaxy with mass $\gtrsim10^{11}\, M_\odot$ at $z\gtrsim1.5$, while the source (sub-)mm continuum and line emissions are amplified by factors $\mu\sim6-13$. We estimate extremely compact sizes $\lesssim0.5$ kpc for the star-forming region and $\lesssim 1$ kpc for the gas component, with no clear evidence of rotation or of ongoing merging events. We perform broadband SED-fitting and retrieve the intrinsic de-magnified physical properties of the source, which is found to feature a very high star-formation rate $\gtrsim10^3\, M_\odot$ yr$^{-1}$, that given the compact sizes is on the verge of the Eddington limit for starbursts; the radio luminosity at 6 cm from available EVLA observations is consistent with the star-formation activity. The galaxy is found to be extremely rich in gas $\sim10^{11}\, M_\odot$ and dust $\gtrsim10^9\, M_\odot$. The stellar content $\lesssim10^{11}\, M_\odot$ places the source well above the main sequence of starforming galaxies, indicating that the starburst is rather young with estimated age $\sim10^8$ yr.
Our results indicate that the overall properties of HATLASJ113526.2-01460 are consistently explained by in-situ galaxy formation and evolution scenarios.
\end{abstract}




\section{Introduction} \label{sec:intro}

Sub-millimetre galaxies (SMGs) are the main protagonists of the star formation at early cosmic times (\citealt{Blain1996}, \citealt{Casey2014}).  
It is well established, that a substantial contribution at the peak of the cosmic Star Formation Rate (SFR) density comes from these heavily dust-obscured objects, featuring a sub-millimeter (sub-mm) flux density $S_{870 \mu m}\gtrsim 1$ mJy and extremely high SFRs, up to $\sim 10^3$ M$_{\odot}$ $\hbox{yr}^{-1}$ (e.g. \citealt{Simpson2020}, \citealt{Dudzeviciute2020}). 
Because of their huge dust content these objects are heavily obscured in optical bands and extremely bright in far-infrared (FIR)/sub-mm bands where the light of newborn stars, reprocessed by dust, is re-emitted.
Moreover, SMGs have been identified as the progenitors of massive quiescent early-type galaxies and constitute the ideal laboratories to test galaxy evolutionary models.  
For example, in in-situ co-evolutionary scenarios (\citealt{Lapi2014,Lapi2018},  \citealt{Pantoni2019}), the intense star formation activity is accompanied by an exponential growth of the active nucleus, whose feedback will eventually sweep away the interstellar medium. The star formation is thus stopped on a relatively short timescale while the nucleus shines as an optical quasar. 

In the last years, an even more extreme population of heavily obscured SMGs has been discovered. These objects are missed in optical/near-IR (NIR) surveys and have been found up to very high redshifts (z$\sim$ 6; \citealt{Riechers2013,Riechers2020}, \citealt{Marrone2018}). These heavily obscured star-forming galaxies often lack of a counterpart even in deep NIR observed-frame Hubble Space Telescope (HST) (\citealt{Wang2019}, \citealt{Gruppioni2020}) or either show extreme red colors ($H-3.6$ $\mu$m $>4$; see e.g. \citealt{Wang2016}) and are visible only from observed-frame mid-IR(MIR) images performed e.g. with the \textit{Spitzer}/Infrared Array Camera (IRAC). Samples of optical/NIR dark objects have been detected by observing deep CO line emission (\citealt{Riechers2020}), and have been efficiently selected in sensitive radio observations (\citealt{Talia2021}, \citealt{Enia2022}).
These peculiar objects provide a significant and previously unknown contribution to the cosmic SFR density at z$\gtrsim$3 estimated to be at least 10$\%$ up to $25-40\%$ with respect to the one inferred from UV-selected populations (\citealt{Wang2019}, \citealt{Williams2019}, \citealt{Gruppioni2020}, \citealt{Talia2021}, \citealt{Enia2022}).

The studies conducted so far are however limited by the poor angular resolution and sensitivity in MIR/FIR and sub-mm bands, causing confusion problems and prohibiting a detailed investigation of the physical properties of optical/NIR dark galaxies and the conditions of their Interstellar Medium (ISM). In the last years, Atacama Large Millimeter/submillimeter Array (ALMA) deep field observations strongly improved the quality of the observations of high redshift dusty galaxies, detecting SMGs up to flux density limits of $S_{\rm 870 \mu m} \sim 0.1 - 1$ mJy (\citealt{Aravena2016}, \citealt{Walter2016}, \citealt{Dunlop2017}, \citealt{Franco2018}, \citealt{Hatsukade2018}). However, even high angular resolution studies indicate that these objects are extremely compact, with typical intrinsic sizes of a few tenths of an arcsec (\citealt{Pantoni2021}, \citealt{Massardi2017}), hence very hard to resolve.

Gravitational lensing enables the observation of regions in the luminosity-redshift space of these sources, that would be otherwise unattainable with current instrumentation in reasonable integration times. The gravitational magnification of the foreground lens increases the apparent luminosity proportionally to the magnification $\mu$ and stretches the angular sizes by a factor $\sqrt{\mu}$. This behavior offers the unique possibility of studying down to sub-kpc scales the properties of objects otherwise not exceptionally bright, massive, or peculiar, and belonging to the dusty star-forming galaxy population bulk at the peak of cosmic star formation.
Several works demonstrated the effectiveness of sub-mm surveys in selecting strong lensing events adopting a flux density threshold of 100 mJy at 500 $\mu$m, in correspondence of a steep drop in the number counts of dusty star-forming galaxies at sub-mm wavelengths (\citealt{Blain1996}, \citealt{Negrello2010}, \citealt{Lapi2012}) where, thanks to the magnification, they emerge as the bright tail of the population count distribution, thus minimizing the probability of possible contaminants, such as flat spectrum radio sources and low redshift spiral galaxies.

Moreover, in the FIR/sub-mm regime, while the high$-z$ lensed dusty star-forming galaxies are particularly bright, negligible signal comes from the foreground lenses, which are often massive ellipticals at $z<1$ that dominate the signal in optical bands.
Several surveys conducted in the last decade with the \textit{Herschel} Space Observatory led to the discovery of numerous strong lensing events. The \textit{Herschel} Multi-tiered Extragalactic Survey (HerMES; \citealt{Oliver2012}) identified 11 lensed galaxies over 95 deg$^2$ (\citealt{Wardlow2013}); \cite{Nayyeri2016} selected other 77 candidate lensed galaxies in the HerMES Large Mode survey (HeLMS; \citealt{Oliver2012}) and in the \textit{Herschel} Stripe 82 Survey (HerS; \citealt{Viero2014}). In particular, the \textit{Herschel} Astrophysical Terahertz Large Area Survey (H-ATLAS; \citealt{Eales2010}) is the widest area (600 deg$^2$) extragalactic survey undertaken with \textit{Herschel} and has provided a sample of more than a hundred thousands dusty star-forming galaxies at high redshift.
Among the H-ATLAS survey, a sample of 80 candidate strongly lensed dusty star-forming galaxies has been selected in \cite{Negrello2017} by means of a simple flux density selection ($S_{500\mu\rm m}>100\,$mJy). Only 21 of them have been confirmed to be lensed thus far. Recently, another sample of 11 candidates has been selected by \cite{Ward2022} in the H-ATLAS Third Data Release conducted in the South Galactic Pole (SGP).
The recent work of \cite{Shu2022} exploited lensing effects generated from galaxy clusters in order to systematically search for optically dark galaxies. Follow-ups performed with JCMT/SCUBA ($\sim$ 850\,$\mu$m) and ALMA ($\sim$ 870\,$\mu$m) for their sample reach a flux limits $\sim$ 3 times deeper than blank fields, highlighting the capabilities of gravitational lensing in detecting even more hidden and dark objects.

In this work, we present the lens modeling, the source reconstruction, and Spectral Energy Distribution (SED) analysis of HATLASJ113526.3$-$014605 (J1135 hereafter), also called G12v2.43 or G12H43, an optical/NIR dark strongly lensed galaxy at  $z = 3.1276$ belonging to the \cite{Negrello2017} lensed candidate sample, featuring a flux density at 500 $\mu$m amounting to $204\pm 8.6$ mJy. The plan of the paper is the following: in Sect. \ref{sec:target} we present the target of our analysis, and describe the archival ALMA observations and the available ancillary data in other bands; Sect. \ref{sec:lens_model} and \ref{sec:SED fitting} describe respectively the lens modeling and source reconstruction and the SED fitting analysis; finally, we discuss and summarise our results in Sect. \ref{sec:discussion} and \ref{sec:conclusions}. 
Throughout the work, we adopt the standard flat $\Lambda$CDM cosmology \citep{Plank2020} with rounded parameter values: matter density $\Omega_M = 0.32$, dark energy density $\Omega_{\Lambda} = 0.63$, baryon density $\Omega_b = 0.05$, Hubble constant $H_0=100h $ km$\hbox{s}^{-1}$Mpc$^{-1}$ with $h = 0.67$, and mass variance $\sigma_8 = 0.81$ on a scale of 8 $h^{-1}$ Mpc. At the redshift of the source 1 arcsec corresponds to 7.8\_kpc.

\section{The target}\label{sec:target}

J1135 is part of the sample of 80 (candidate) strongly lensed galaxies (\citealt{Negrello2017}) located in the equatorial GAMA 12$^{\rm th}$ field (RA=11:35:26, dec=-01:46:07, J2000).
The spectroscopic redshift of $z = 3.1276$ of the background lensed source was obtained from blind CO searches with the Zpectrometer ultrawideband spectrometer on the Green Bank Telescope (GBT) (\citealt{Harris2012}) and confirmed by the Northern Extended Millimeter Array (NOEMA) observations (\citealt{Yang2017}). So far, no redshift measurement is available for the foreground lens. \cite{Andreani2018} presented observations of high CO transition (J=7-6) obtained with the Atacama Pathfinder EXperiment (APEX)/SEPIA band 5 receiver for the background object. From the comparison of the CO(7-6) transition with the CO(1-0) and CI(2-1) the authors pointed out to the presence of a large excitation status in the ISM of J1135.

Moreover, \cite{Vishwas2018} reported bright [OIII] 88 $\mu$m emission for J1135 detected through the z Early Universe Spectrometer (ZEUS-2) on APEX attributed to ionized hydrogen regions around massive stars. From the SED-fitting of the multi-band photometry of J1135, the authors predicted J1135 to be a young, gas-rich starburst galaxy.

The object has also been targeted by Submillimeter Array (SMA) high spatial resolution ($\sim$ 0.8 arcsec) observations described in \cite{Bussmann2013a}, but only marginally resolved. For this reason, its lensed nature has been debated in the works described above. 


\subsection{ALMA observations}\label{sec:alma_obs}
The object is part of low ($\lesssim$ 2 arcsec) resolution observations in band 3 (2017.1.01694.S, PI: Oteo) aimed at tracing dense molecular gas through J=4-3 transitions of HCN, HCO+, and HNC molecules. J1135 was also included in a project (2019.1.00663.S, PI: Butler) whose main goal was to investigate outflows in high redshift star-forming galaxies by tracing OH+ and CO(9-8) lines.

In the following, we describe the calibration, imaging and analysis of further data sets with the highest angular resolution available in the ALMA Science Archive for J1135. 
These spatially resolved ALMA follow-ups reveal an almost complete Einstein ring, confirming out of any doubt the lensed nature of this system. 

The object has been target of ALMA Cycle 4 high-resolution follow-up in band 8 (2016.1.01371.S, PI: Amit) aimed at resolving the lensed morphology of the source and tracing the continuum at $\sim0.7$ mm and the C[II] 158 $\mu$m FIR line. 
The continuum was observed exploiting four base-bands of width 1.98 GHz, centred at 472.284, 470.451, 460.409, 458.534 GHz and  composed by 128 channels each. 

\begin{figure*}
    \centering
    \includegraphics[width=5.2cm]{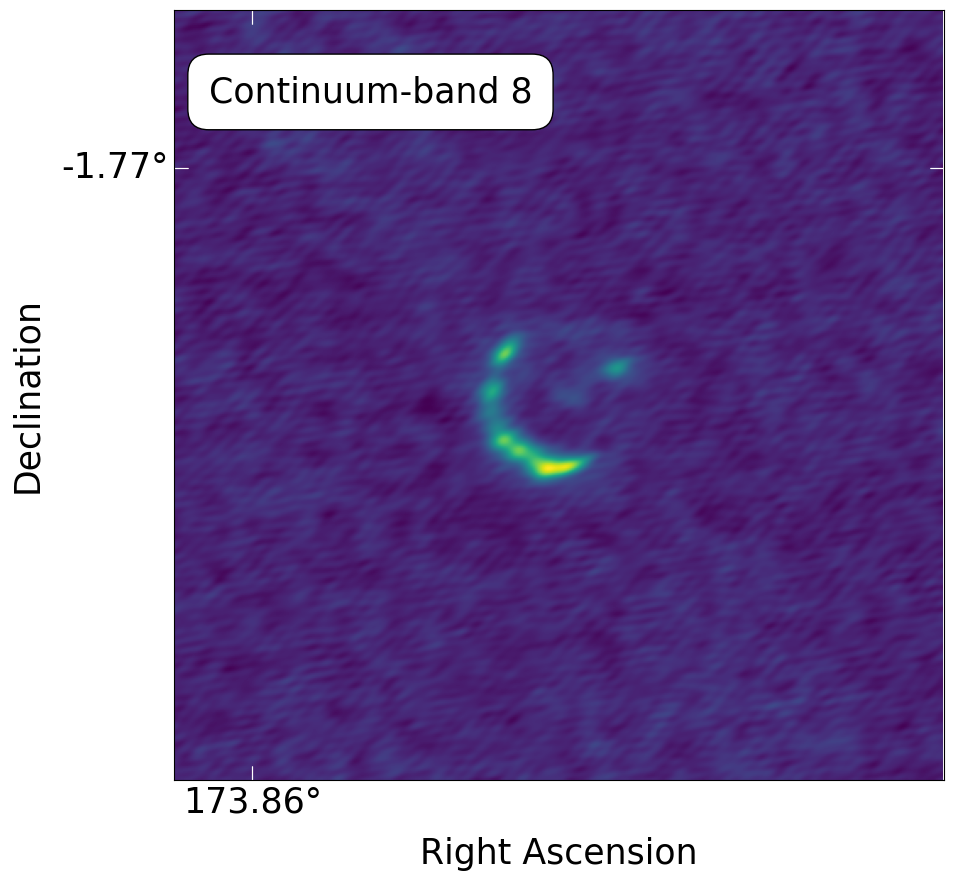}
    \includegraphics[width=4.3cm]{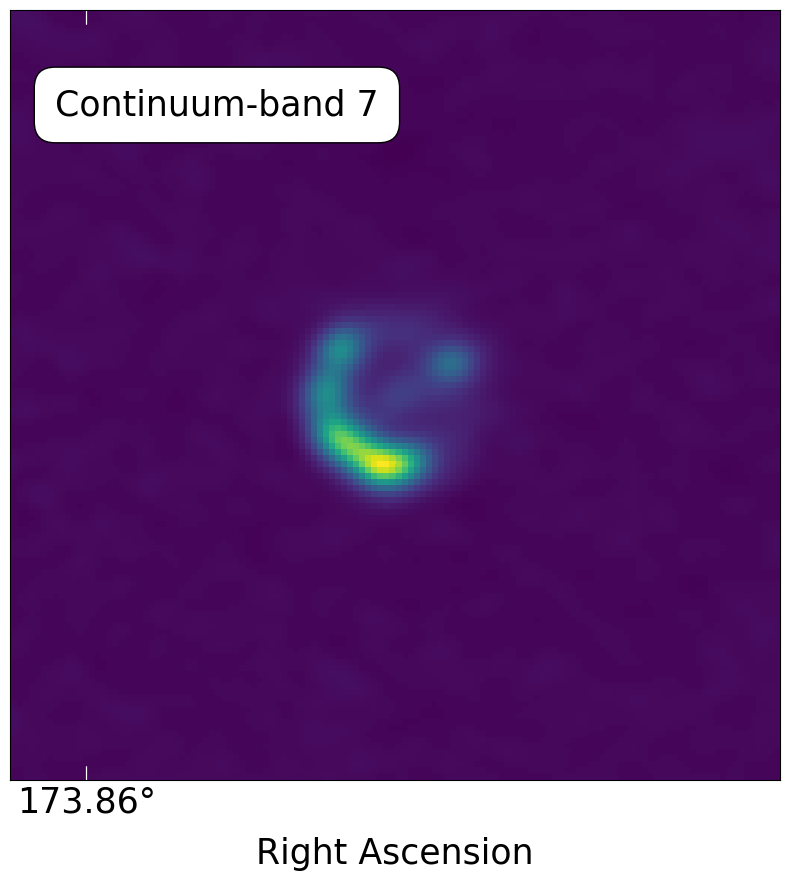}
    \includegraphics[width=4.3cm]{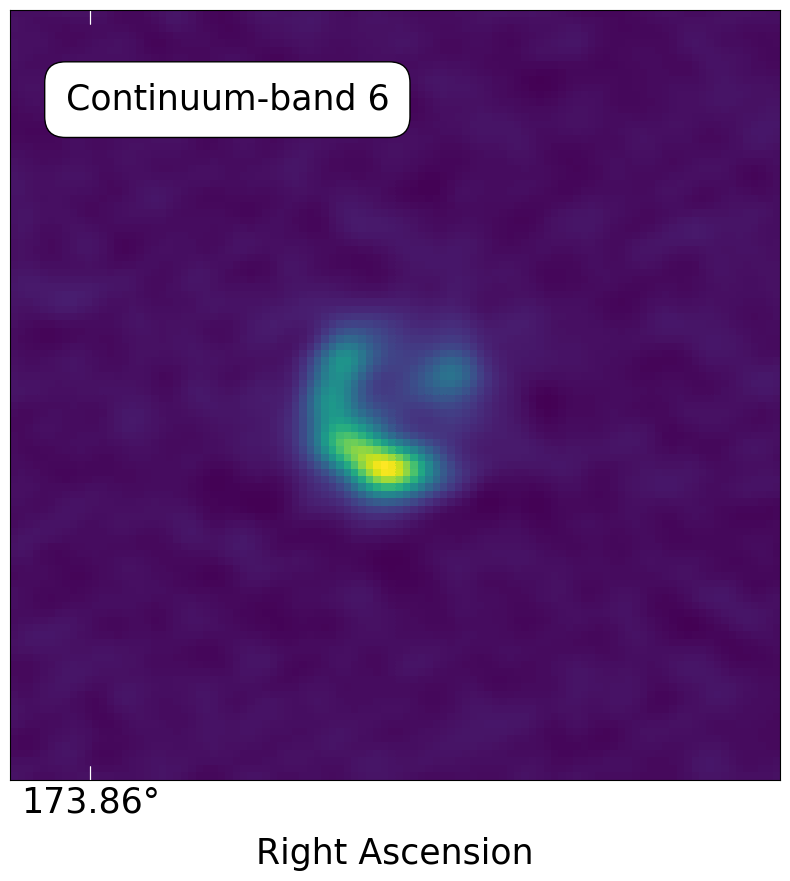}

    \caption{ALMA band 8, 7 and 6 continuum emission for J1135.}   \label{fig:continuum}
\end{figure*}

\begin{figure*}
    \centering
    \includegraphics[width=4.3cm]{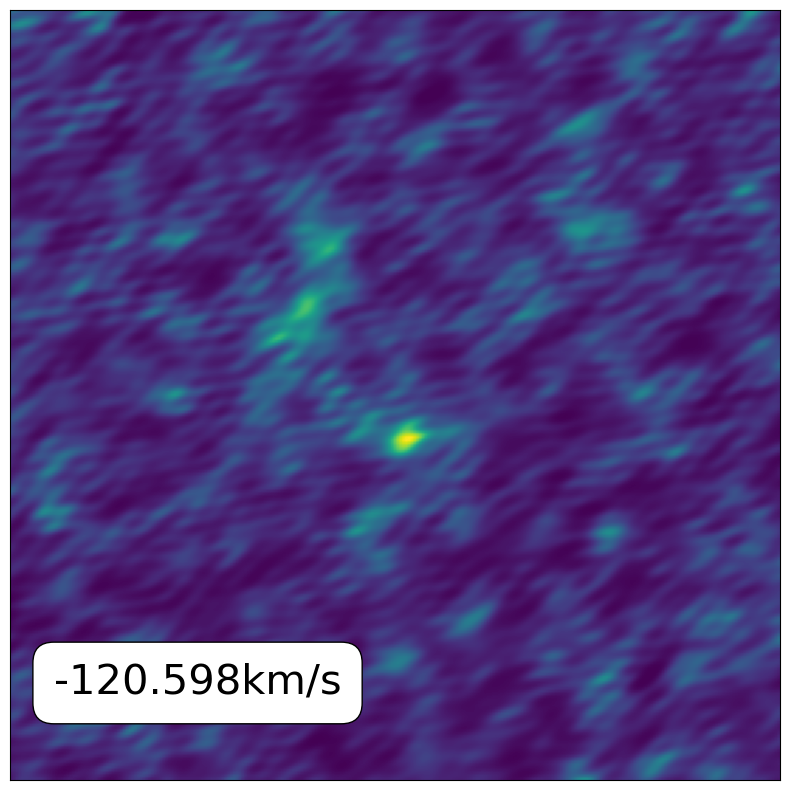}
    \includegraphics[width=4.3cm]{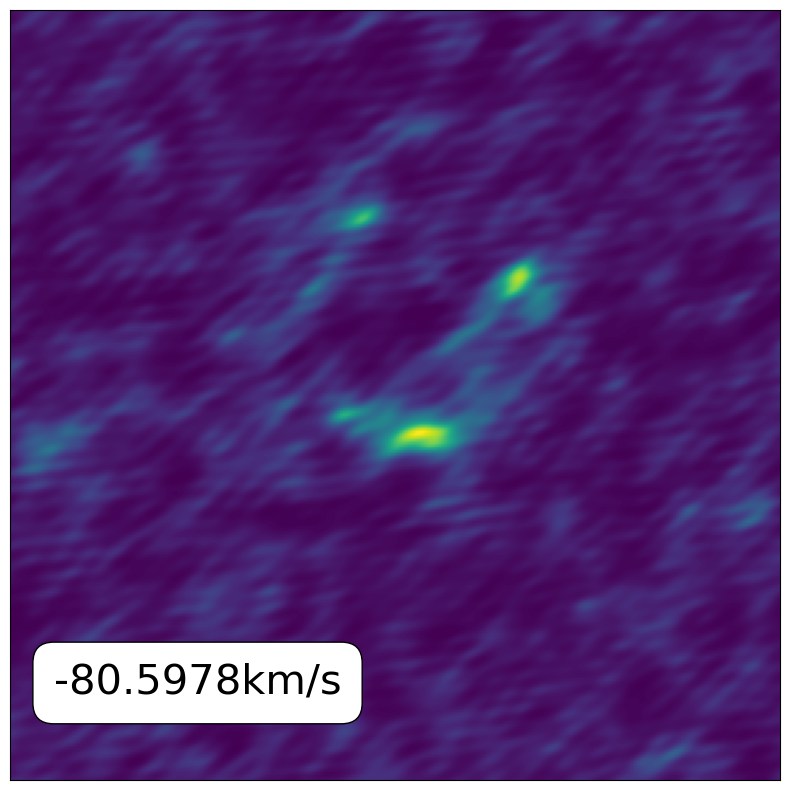}
    \includegraphics[width=4.3cm]{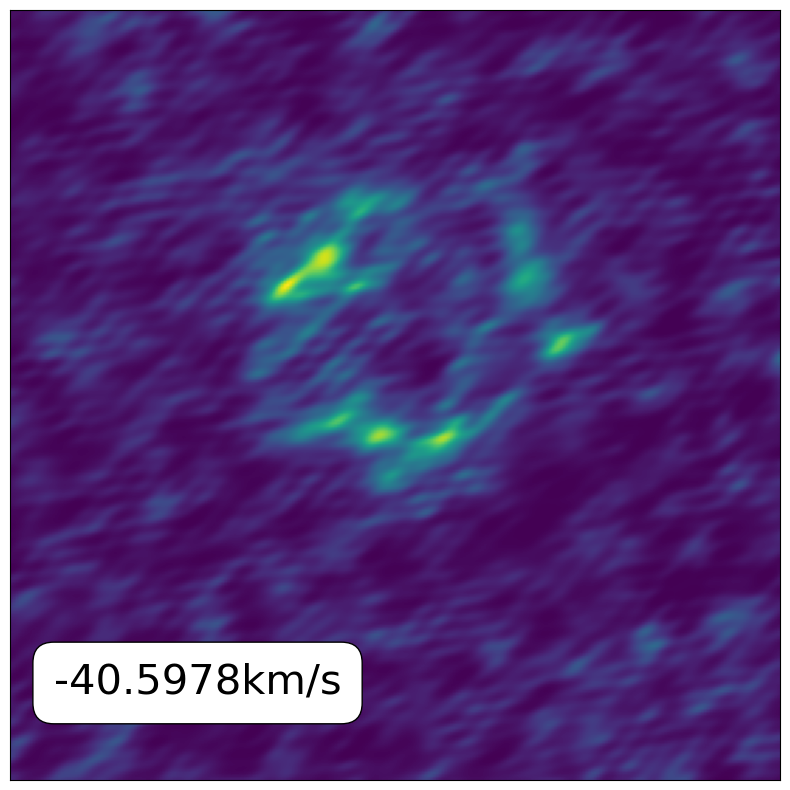}
    \includegraphics[width=4.3cm]{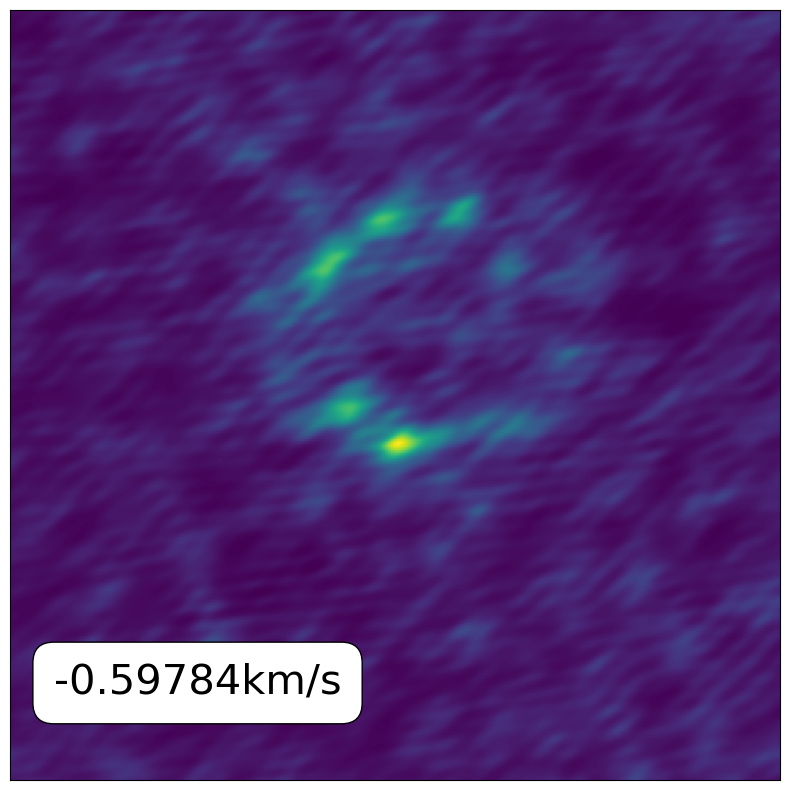}
     \includegraphics[width=4.3cm]{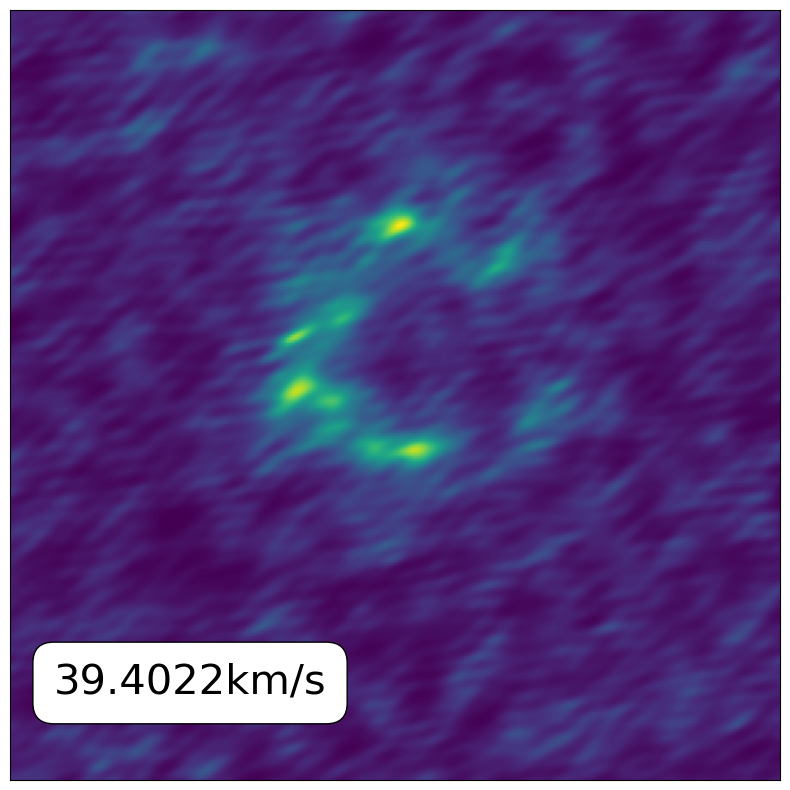}
      \includegraphics[width=4.3cm]{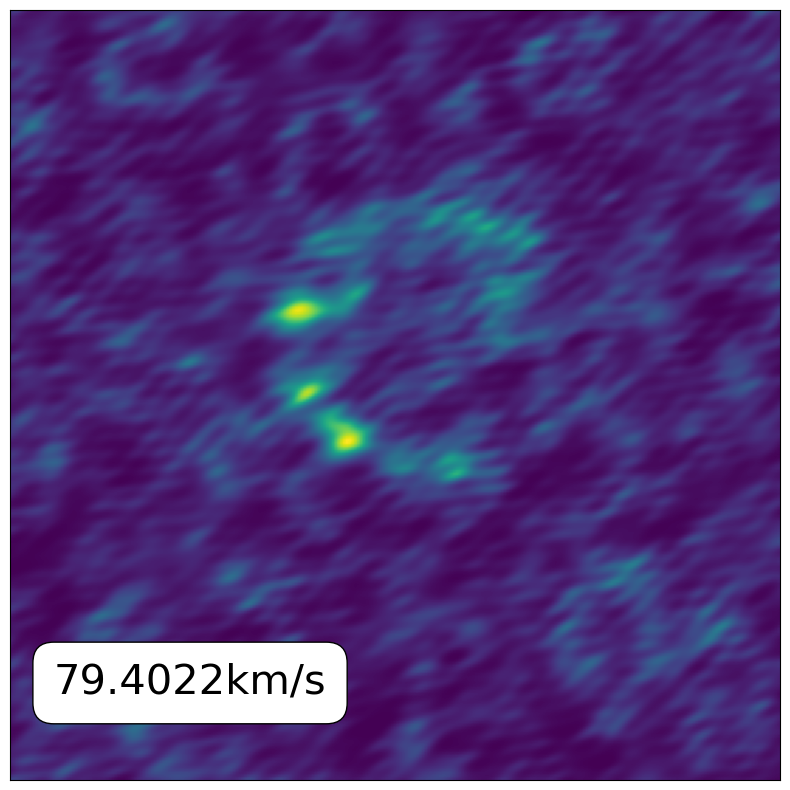}
       \includegraphics[width=4.3cm]{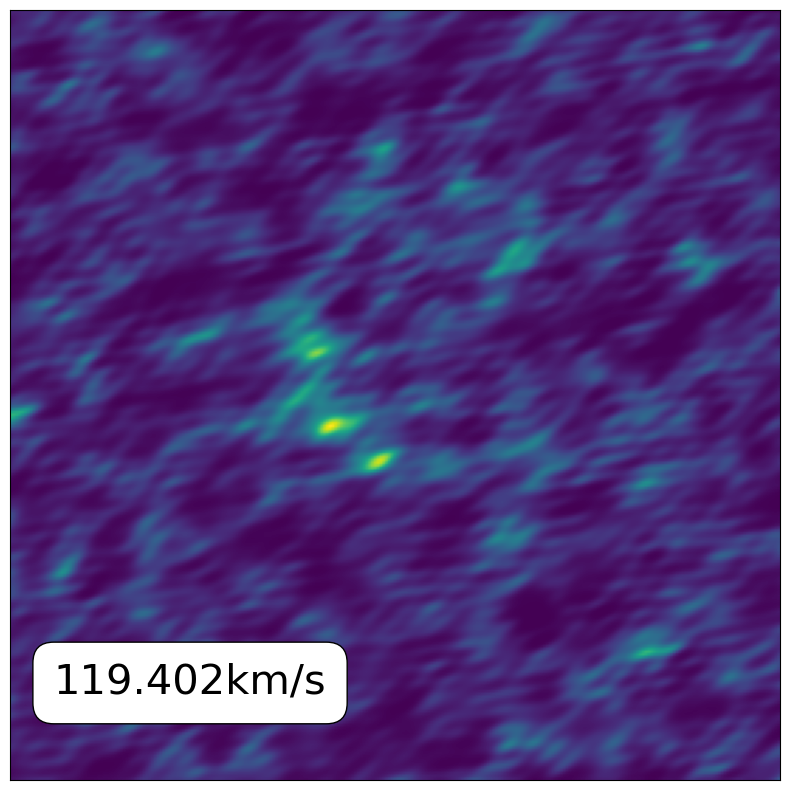}
    \caption{Channel maps and spectrum of the CII line emission for J1135.}   \label{fig:cII}
\end{figure*}

We re-calibrated the raw data using the Common Astronomy Software Applications (CASA) package version 4.7.2 and running the provided calibration scripts. The continuum subtraction was manually done using the task uvcontsub. 
Imaging has been performed manually adopting a Briggs weighting scheme, which assumes a robustness factor of 0.5. The properties of the generated images are reported in Table \ref{tab:alma_obs_prop}, the continuum cleaned images are shown in Fig. \ref{fig:continuum}, and Fig. \ref{fig:cII} reports the C[II] channel maps rebinned in a 20 km $\hbox{s}^{-1}$ interval.

The second data-set we examine is part of the ALMA Cycle 6 project (2018.1.00861.S, PI: Yang) carried out with the goal of tracing H$_2$0 and CO (J=8$-$7) lines in candidate lensed galaxies at high redshift ($z\sim 2-4$) in band 6 and 7.
Both observations are performed with the same configuration with a maximum baseline of 1397 m and four spectral windows of 1.875 GHz bandwidth and 240$\times$7.8 MHz channels each.
In Band 6, the H$_2$0(J=2$_{\rm0,2}$-1$_{\rm 1,1}$) and CO(J=8-7) are targeted with two spectral windows respectively centered at 239.376 GHz and 223.583 GHz, while other two windows centered at 235.940 and 221.705 GHz are dedicated to continuum observations.
In Band 7, two spectral windows, centered at 281.766 and 292.621 GHz, target the H$_2$O(J=3$_{\rm2,1}$-3$_{\rm 1,2}$) and H$_2$O(J=4$_{\rm2,2}$-4$_{\rm 1,3}$) lines, while continuum is observed in two windows centered at 280.314 and 294.266 GHz

\begin{figure*}
    \centering
    \includegraphics[width=4.3cm]{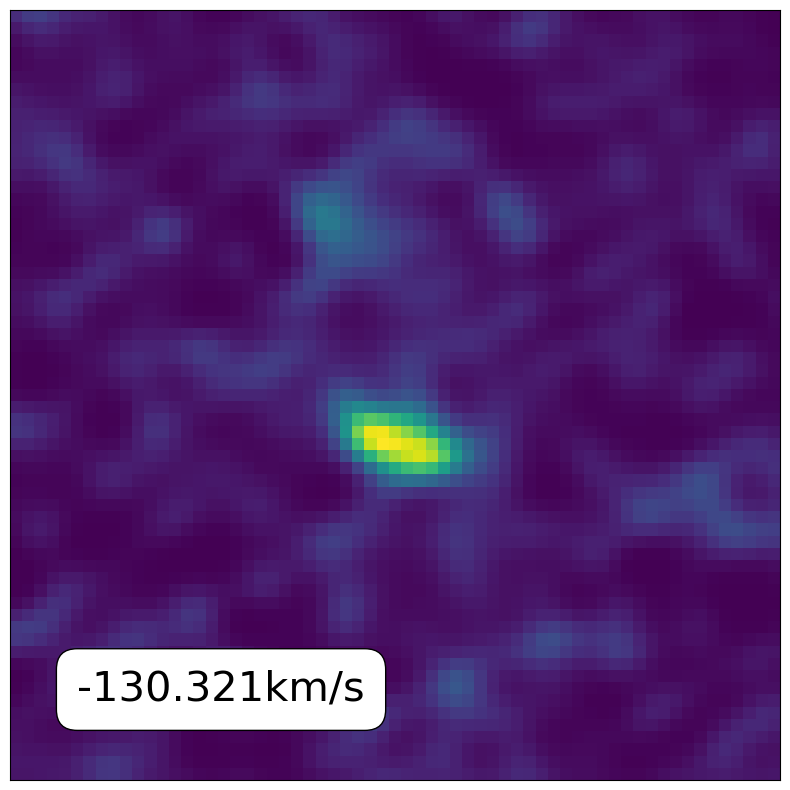}
    \includegraphics[width=4.3cm]{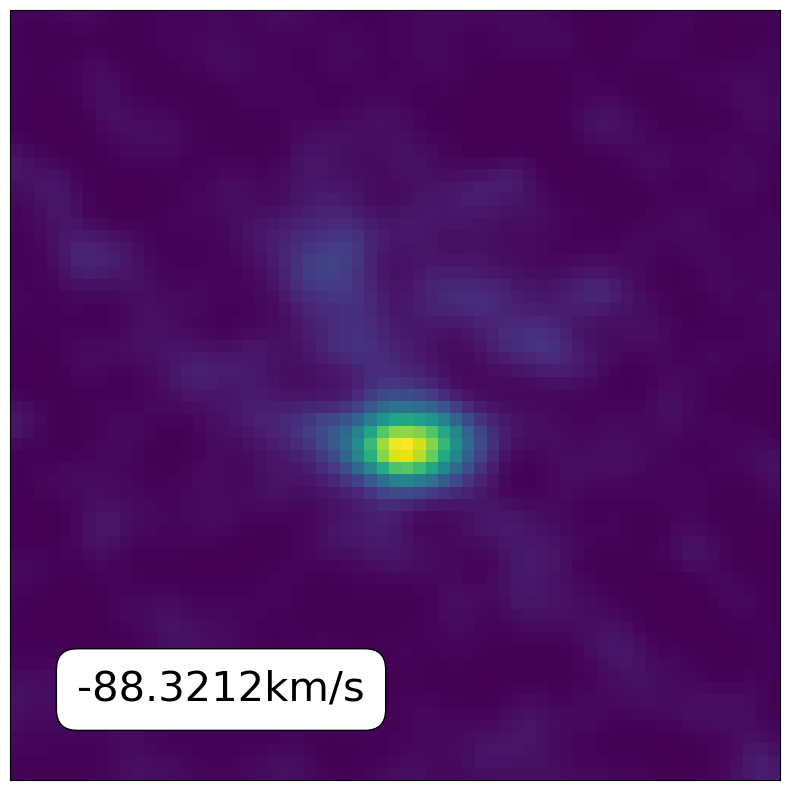}
    \includegraphics[width=4.3cm]{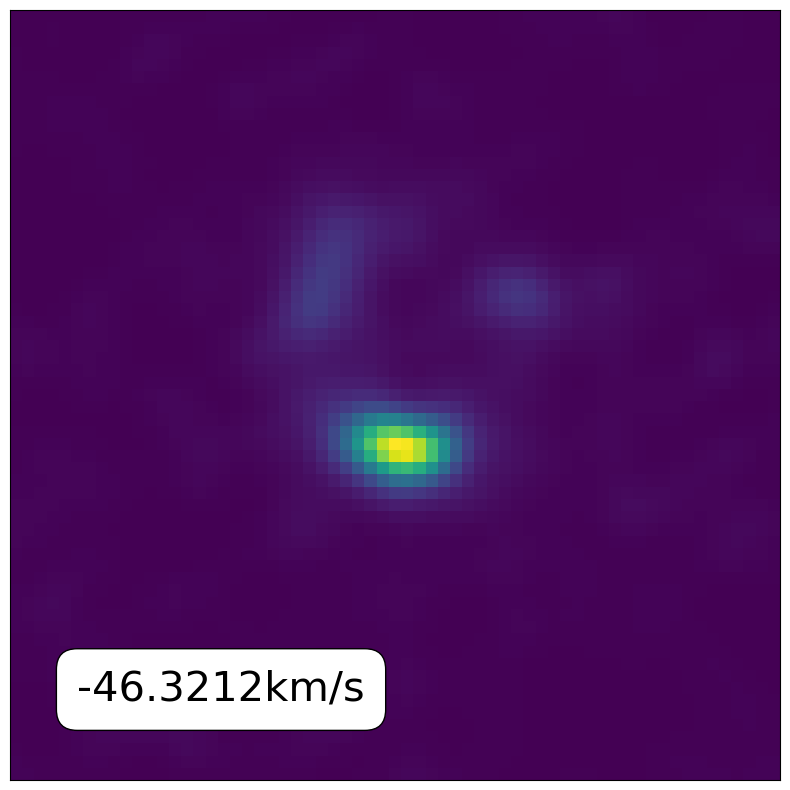}
    \includegraphics[width=4.3cm]{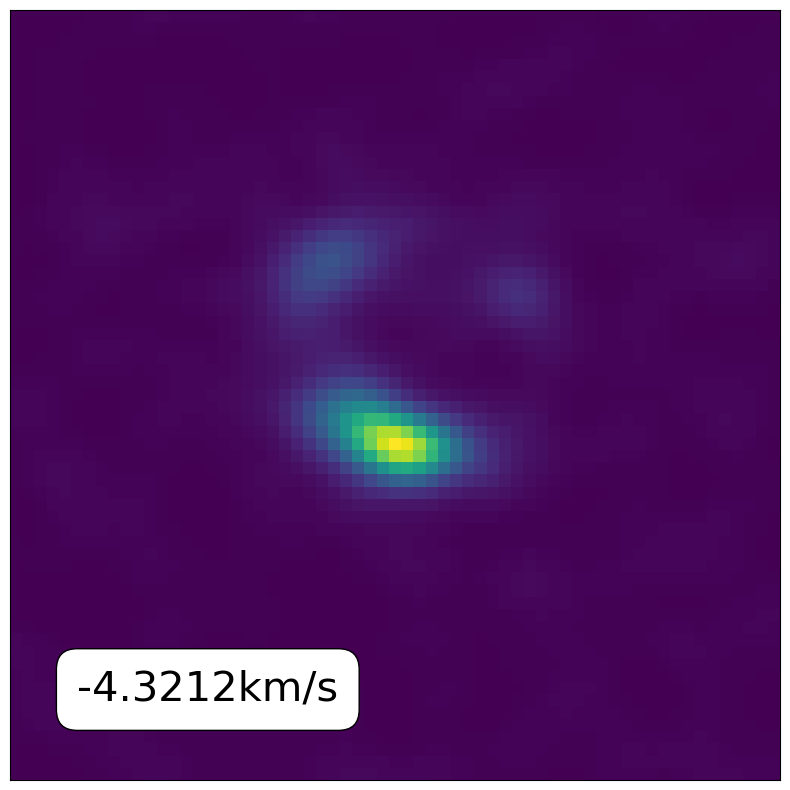}
    \includegraphics[width=4.3cm]{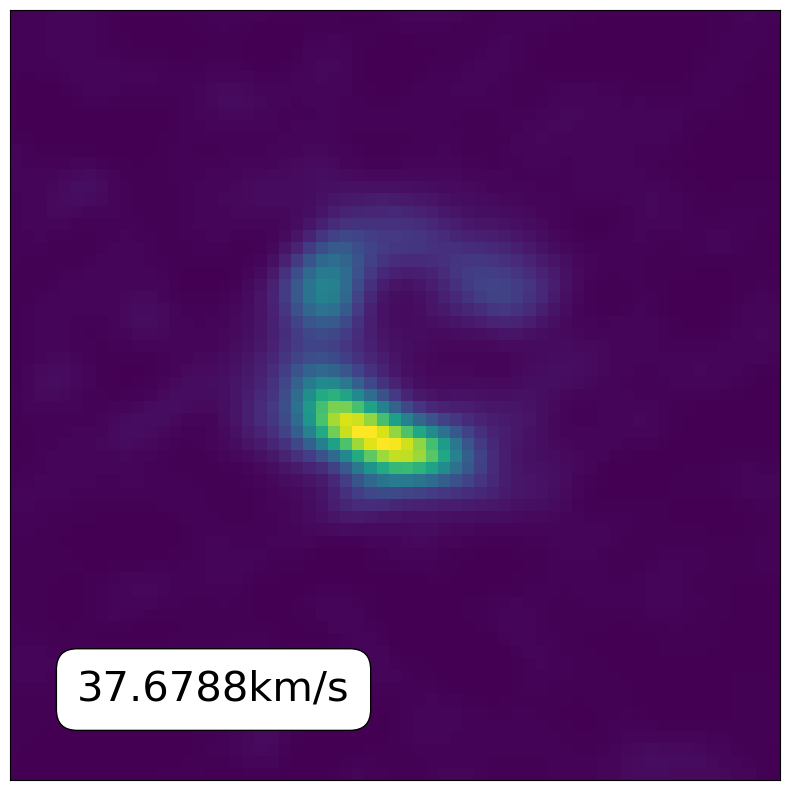}
    \includegraphics[width=4.3cm]{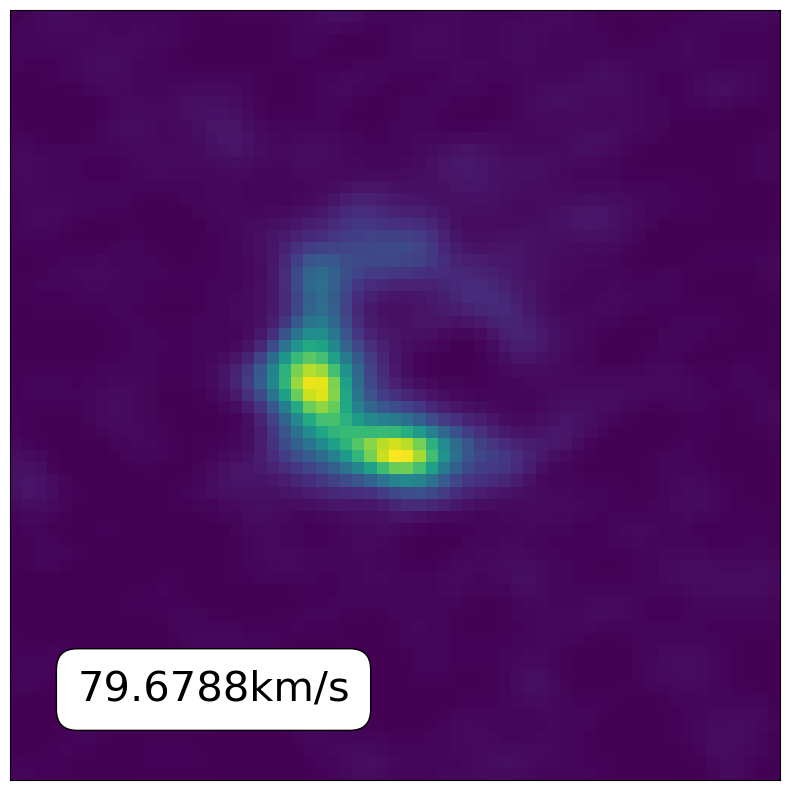}
    \includegraphics[width=4.3cm]{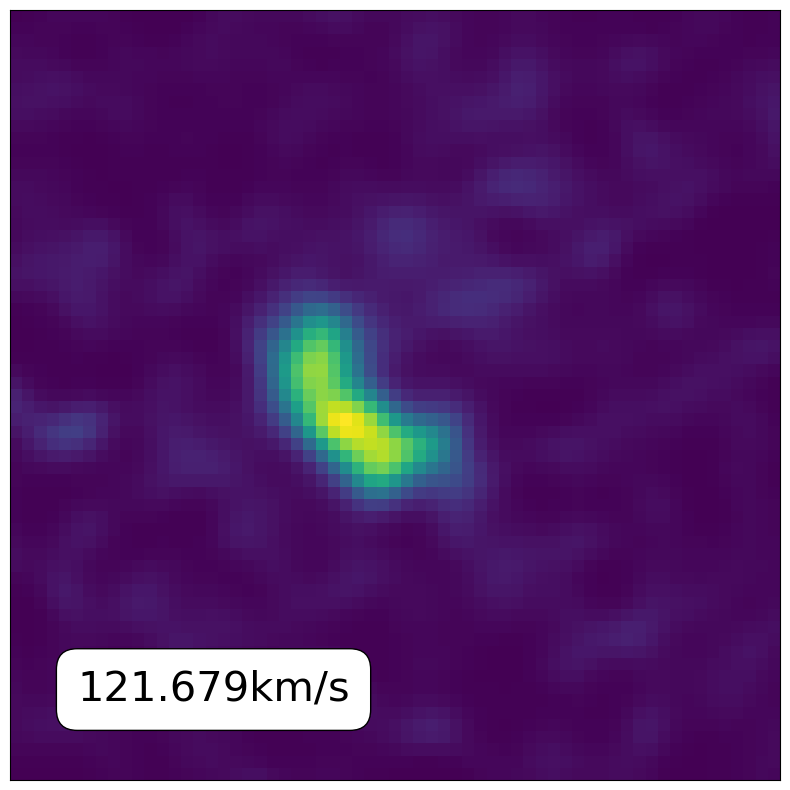}    
    \caption{Channel maps of the CO(8-7) line emission for J1135.}   \label{fig:CO}
\end{figure*}

Calibration is performed running the available pipeline scripts in CASA version 5.4.0-68. Imaging is performed manually adopting a Briggs weighting scheme in both band 6 and 7, with robustness parameter equal to 0.5. 
We image the CO(8-7) line performing an automatic continuum-subtraction.
Fig \ref{fig:CO} reports the CO(8-7) channel maps obtained performing imaging with a channel width of 20 km $\hbox{s}^{-1}$.

The main features of the ALMA data analysed in this work and the properties of the final images are reported in Table \ref{tab:alma_obs_prop}. Note that H$_2$O data cubes included in Cycle 6 observations will not be analysed in this paper.
\vspace{0.5cm}

\subsection{Data analysis}

\begin{table*}
\centering
\caption{Overview of the ALMA observations used in this paper. H$_2$O lines in Cycle 6 data are not analysed in this work. }
\label{tab:alma_obs_prop}
\begin{tabular}{lccc} 
\hline
                              & Cycle 4 B8                            & Cycle 6 B6                           & Cycle 6 B7                            \\ 
\hline
Project ID                    & 2016.1.01371.S                        & 2018.1.00861.S                       & 2018.1.00861.S                        \\
Spectral set-up [MHz]             & \multicolumn{1}{c}{4$\times$128$\times$15.63}                  & \multicolumn{1}{c}{4$\times$240$\times$7.81}                 &  \multicolumn{1}{c}{4$\times$240$\times$7.81}                  \\
Spectral resolution [km$\hbox{s}^{-1}$]          & \multicolumn{1}{c}{10.17}              & \multicolumn{1}{c}{10.48}                & \multicolumn{1}{c}{\ ...}                  \\
Restored beam axes [arcsec$^2$]   & \multicolumn{1}{c}{0.14$\times$ 0.07} & \multicolumn{1}{c}{0.29$\times$0.25} & \multicolumn{1}{c}{0.23$\times$0.2}  \\
Sensitivity [mJy beam$^{-1}$] & 0.541                                 & 0.043                                & 0.025                                 \\
Lines                         & C[II]                                 & CO(8-7)                           & \multicolumn{1}{c}{\ ...}                           \\
\hline
\end{tabular}
\end{table*}

The flux densities derived for the continuum emission of the lensed source are reported in Tab. \ref{tab:photometry}. We also include the flux density value measured from the  archival image of the Band 3 continuum emission mentioned in Sect. \ref{sec:target}. Flux density uncertainties are computed including a 5$\%$ estimation of the flux calibration accuracy:
\begin{equation}
    \delta S_{\rm image} = \sqrt{(\sigma_{\rm image})^2 + (0.05 \times S_{\rm image})^2},
\end{equation}
with $\sigma_{\rm image}$ being the image noise.

\begin{figure*}
    \centering
     \begin{tabular}{c c}
        \textbf{C[II]} & \textbf{CO(8-7)} \\
        \includegraphics[width=8.3cm]{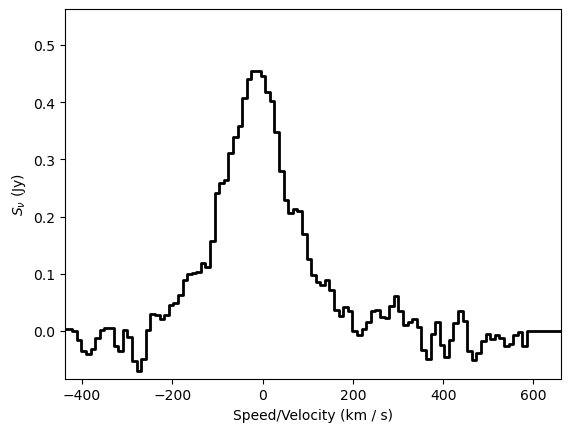} & \includegraphics[width=8.3cm]{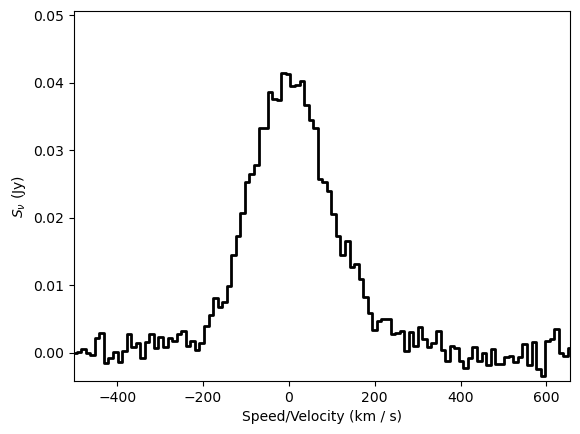}
    \end{tabular}

    \caption{Spectral emission of the C[II] and CO(8-7) lines.}   \label{fig:lines}
\end{figure*}

By fitting the resolved ALMA spectral lines with a single Gaussian profile we obtain the full-width half maximum (FWHM) values for both CO(8-7) and C[II] lines corresponding to $215.\pm4$ and $181\pm5 $ km $\hbox{s}^{-1}$ respectively, in concordance to what is found by GBT and NOEMA CO and H$_2$0 lines analysed in \cite{Harris2012} and \cite{Yang2017}. The peak is detected at $\nu_{\rm obs}=460.504\pm0.003$ GHz for C[II] and $\nu_{\rm obs}=223.356\pm0.0011$ GHz for the CO(8-7), confirming the redshift estimate by \cite{Harris2012} of 3.127 whose associated uncertainties are $\delta z_{\rm C[II]}=\pm 0.005$ and $\delta z_{\rm CO(8-7)}=\pm0.003$ respectively. 
The observed magnified line profiles measured within a region containing the whole source emission are shown in Fig.\ref{fig:lines}.
Following \cite{Carilli2013} we compute the observed magnified C[II] and CO(8-7) luminosities expressed in units of K km s$^{-1}$ as:

\begin{equation}
    L'_{\rm line} = 3.25 \times 10^7 \times S_{\rm line} \Delta v \frac{D_L^2}{(1+z)^3 \nu^2_{\rm obs}}.
\end{equation}

Where $S_{\rm line} \Delta v$ is the measured flux of the line profile (in units of $\hbox{Jy}$ $\hbox{km}$ $\hbox{s}^{-1}$) and $D_L$ is the luminosity distance.
The luminosities expressed in $L_{\odot}$ are computed as $L_{\rm line} = 3 \times 10^{-11} \nu^3_{\rm rest} L'_{\rm line}$. 
The final values computed the C[II] and the CO(8-7) lines are summarised in Table \ref{tab:line_values}.


\begin{table*}
\centering
\caption{Properties of the C[II] and CO(8-7) lines. From the left: measured flux from a single Gaussian profile fit, line luminosity expressed in $L_{\odot}$ and line luminosity expressed in K km s$^{-1} $pc$^2$}
\label{tab:line_values}
\begin{tabular}{lccc} 
\hline
Line    & \begin{tabular}[c]{@{}c@{}}$\mu \Delta v$S$_{\rm line}$\\{[}Jy km s$^{-1}$]\end{tabular} & \begin{tabular}[c]{@{}c@{}}$\mu$L\\{[}$10^{9}$L$_{\odot}$]\end{tabular} & \begin{tabular}[c]{@{}c@{}}$\mu $L'\\{[}$10^{11} $K km s$^{-1} $pc$^2$]\end{tabular}  \\ 
\hline
C[II]   & 82.5$\pm$2.1 & 27.7$\pm$0.7    & 1.35$\pm$0.03   \\              
CO(8-7) & 9.4$\pm$0.1  & 1.54$\pm$0.03 & 0.65$\pm$0.01                                                                        \\
\hline
\end{tabular}
\end{table*}

\subsection{Other bands}\label{sec:multiband_obs}

J1135 is covered by several surveys, such as the Kilo-Degree Survey (KiDS, \citealt{deJong2013}) and the Hyper Suprime-Cam Subaru Strategic Program in the UV/optical bands (\citealt{Aihara2018,Aihara2022}), the VISTA Kilo-degree Infrared Galaxy Public Survey (VIKING, \citealt{Edge2013}), and the UK Infrared Deep Sky Survey Large Area Survey (UKIDSS-LAS, \citealt{Lawrence2007}) surveys in the Near-IR (NIR), the Wide-field Infrared Survey Explorer (WISE, \citealt{Wright2010}) in the MIR. PACS and SPIRE FIR observations are reported in HATLAS First and Second Data Release catalogues (\citealt{Valiante2016}, \citealt{Maddox2017a}). Moreover, the source is covered by the  VLA Faint Images of the Radio Sky at Twenty-Centimeters (FIRST, \citealt{Becker1995}) survey in the radio band, where no emission is detected. 

High resolution NIR follow-up observations are available for J1135. The target was observed as part of the Cycle 19 HST/WFC3 snapshot program (PI: Negrello) at a wavelength of $\lambda=1.15$ $\mu$m (see \citealt{Negrello2014} for further details of the observations) and with the Keck telescope in Adaptive Optics (AO) in the Ks band (\citealt{Calanog2014}). No successful detection has been found in the Ks image, while a marginal emission ($\lesssim 3\sigma$) is present in the HST image, however, given the un-sufficient sensitivity and angular resolution it is not possible to unambiguously confirm whether it is associated to the foreground lens or the background source.

The object is also detected in MIR observations available in the \textit{Spitzer}/IRAC Data Archive (PI: Cooray) and described in \cite{Ma2015}, covering IRAC channel 1 and channel 2, at 3.6 $\mu$m and 4.8 $\mu$m respectively.

\begin{figure}
    \centering
 
        \includegraphics[width=0.45\textwidth]{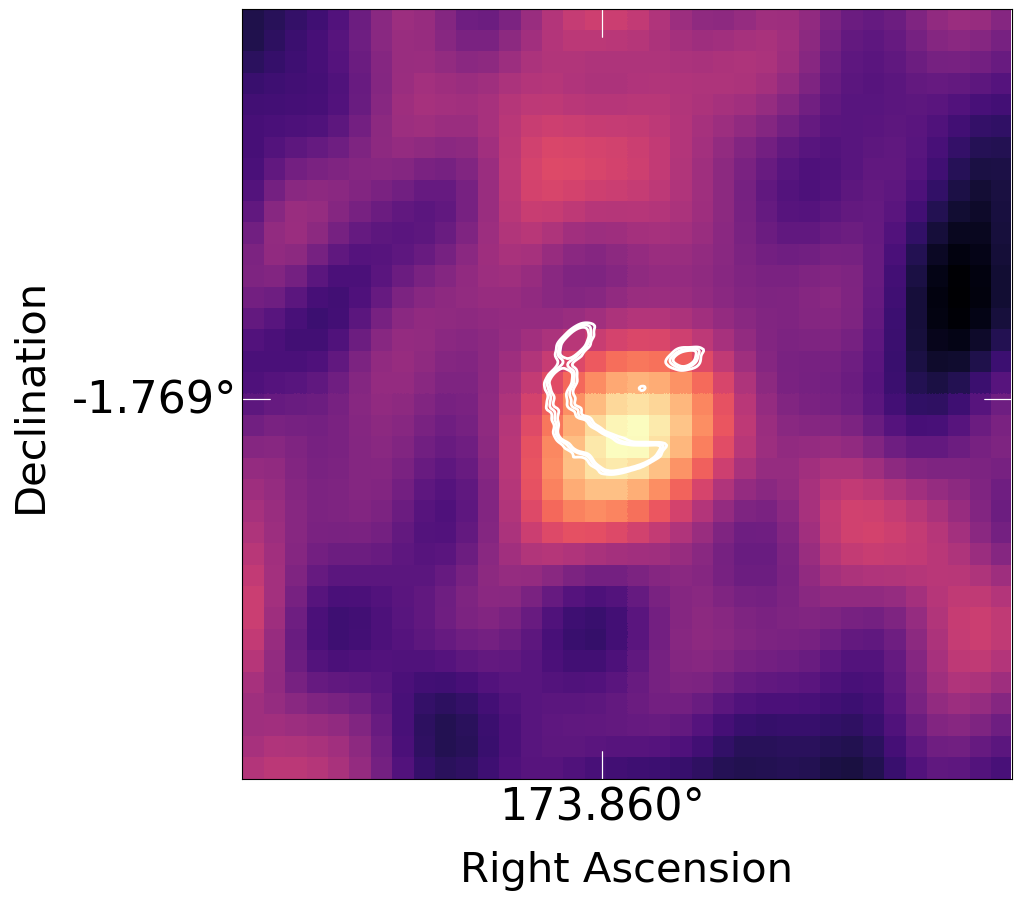} 

    \caption{Image of the EVLA detection at 6 cm for J1135. White contours represent the ALMA continuum emission in band 8 at 9,7,5,3$\sigma$ } \label{fig:EVLA_det}
\end{figure}

In addition, we find EVLA radio data in the NRAO Archive, in particular follow-ups in C-band centred ad $\sim 6$ GHz (project code: 16A-240, PI: Smith). Data are processed by running the calibration scripts, cleaning is performed manually with CASA adopting an interactive mask. The final image reaches a mean rms of $\sim0.013$ mJy beam$^{-1}$ and a restored beam ellipse of 1.13$\times$0.84 arcsec (see Fig. \ref{fig:EVLA_det} ).

\begin{figure*}
    \centering
	\includegraphics[width=4.2cm]{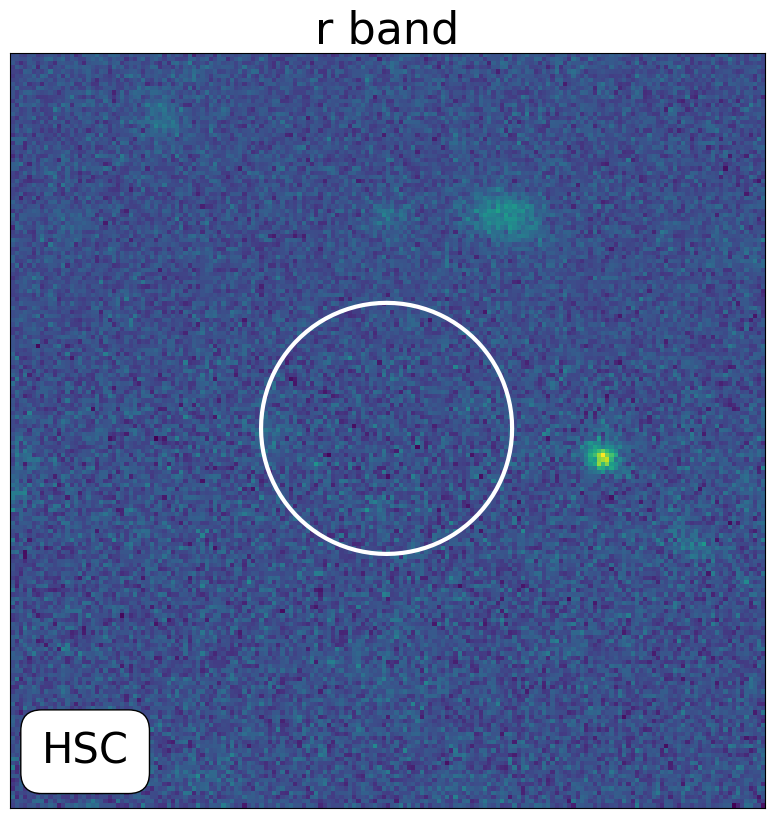}  
	\includegraphics[width=4.2cm]{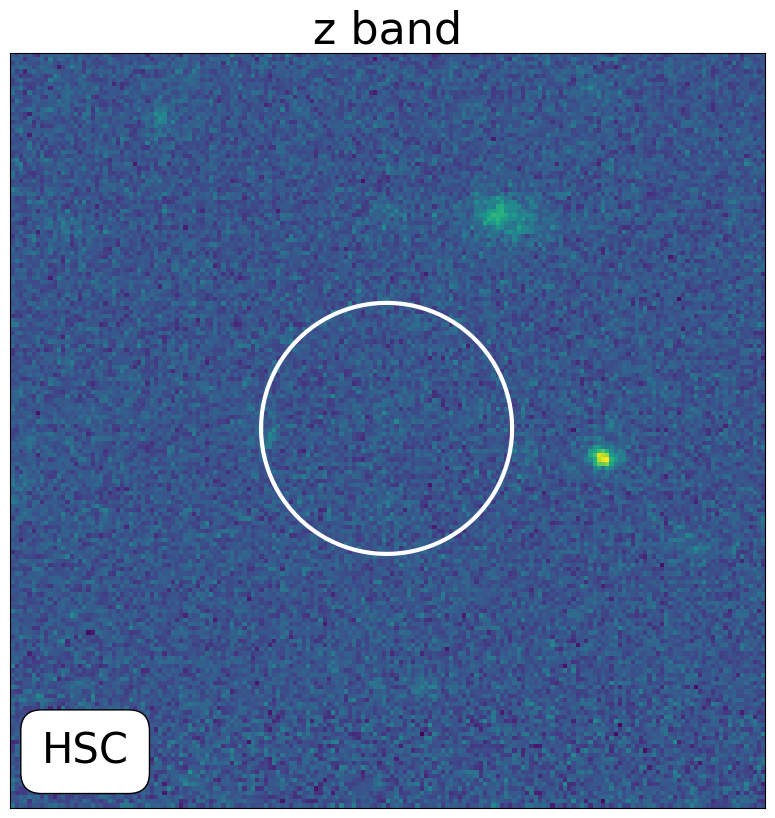} 
	\includegraphics[width=4.2cm]{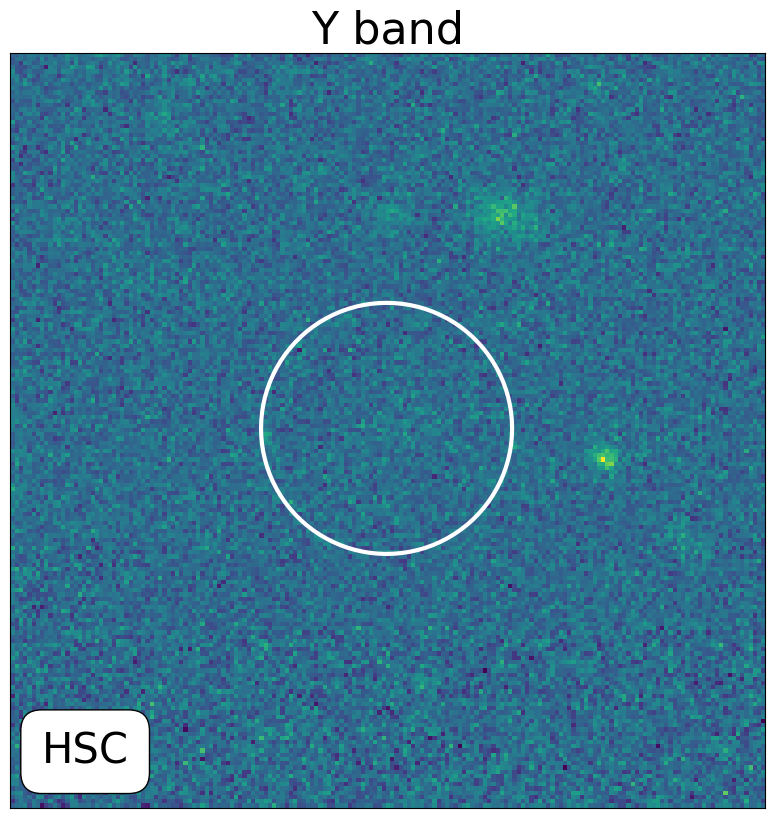} 
	\includegraphics[width=4.2cm]{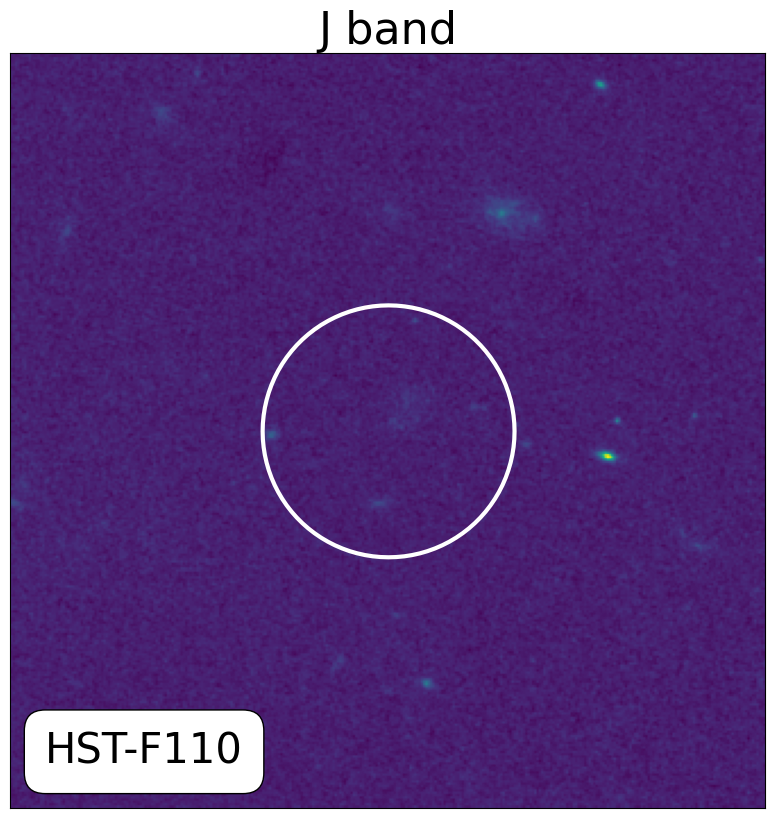} 
	\includegraphics[width=4.2cm]{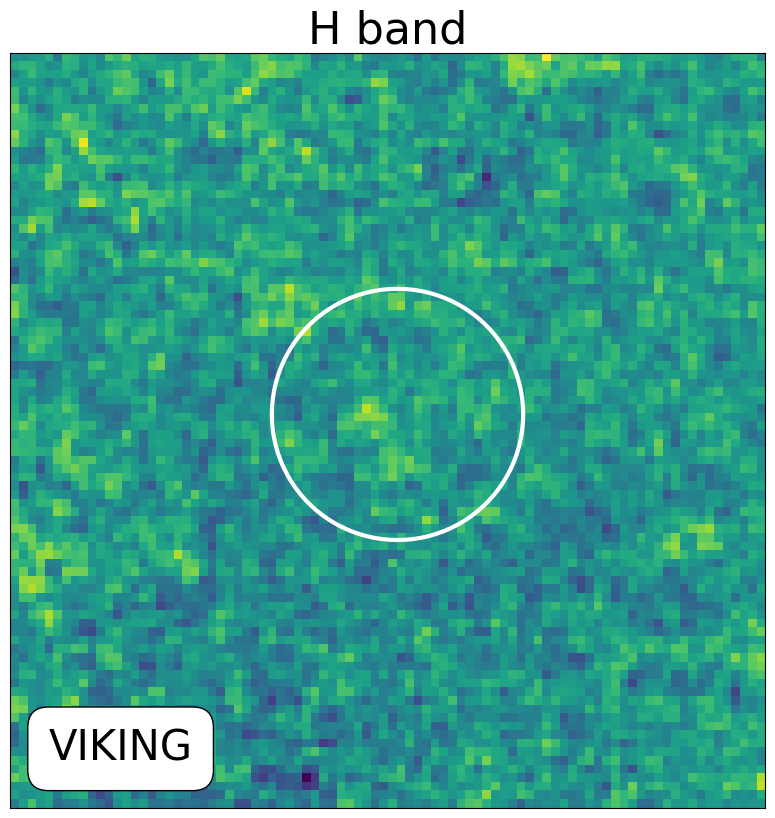} 
	\includegraphics[width=4.2cm]{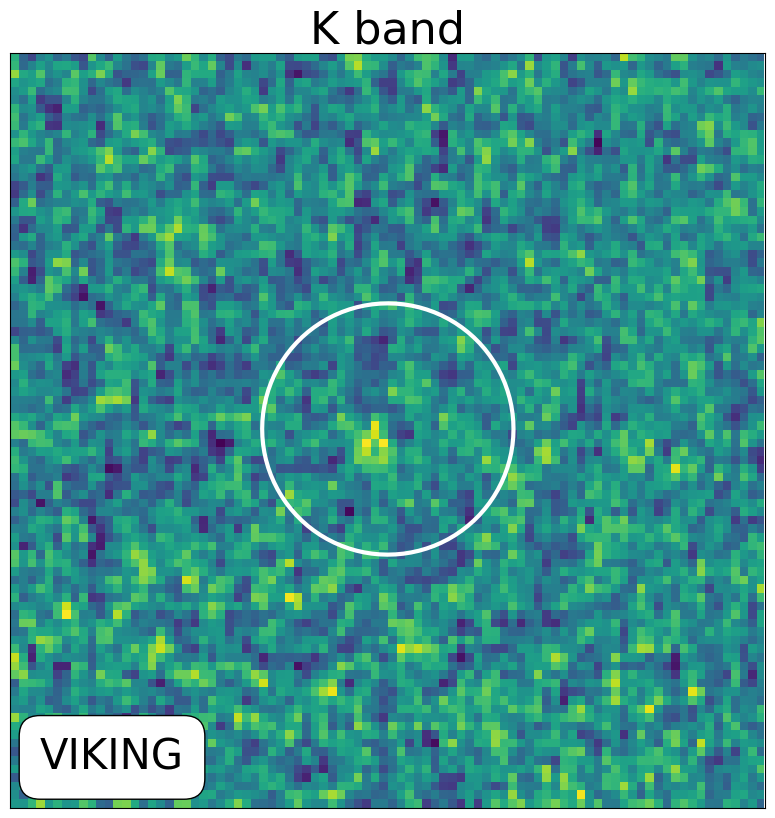} 
	\includegraphics[width=4.2cm]{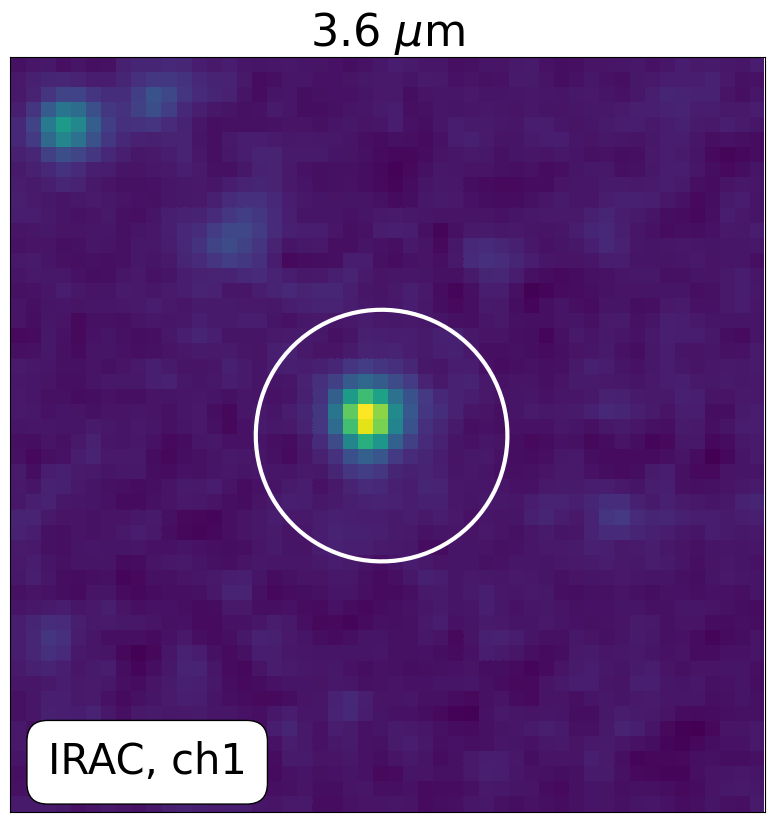}
	\includegraphics[width=4.2cm]{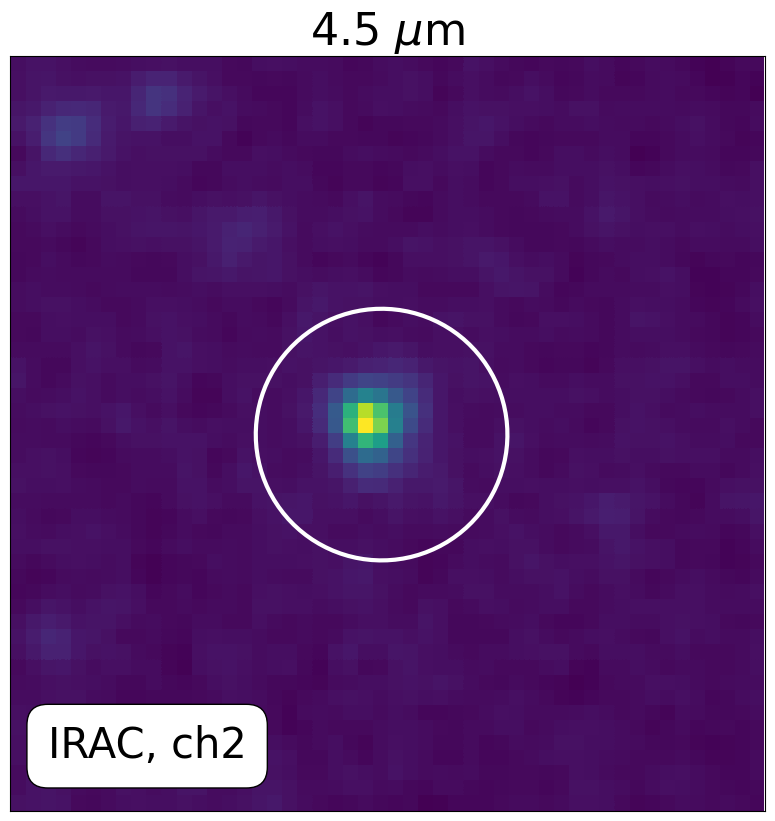}
	\includegraphics[width=4.2cm]{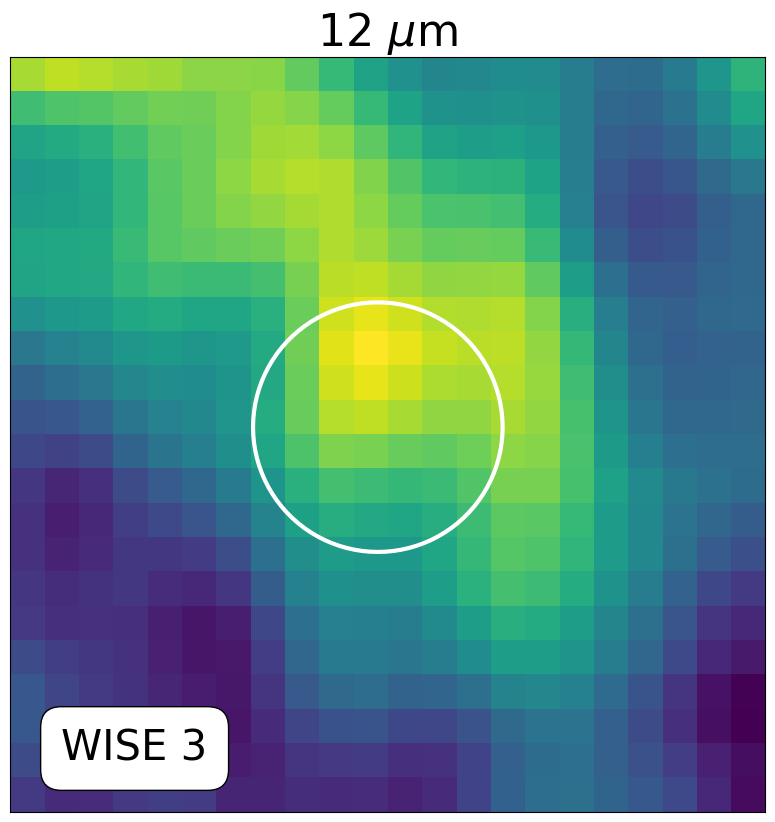} 
	\includegraphics[width=4.2cm]{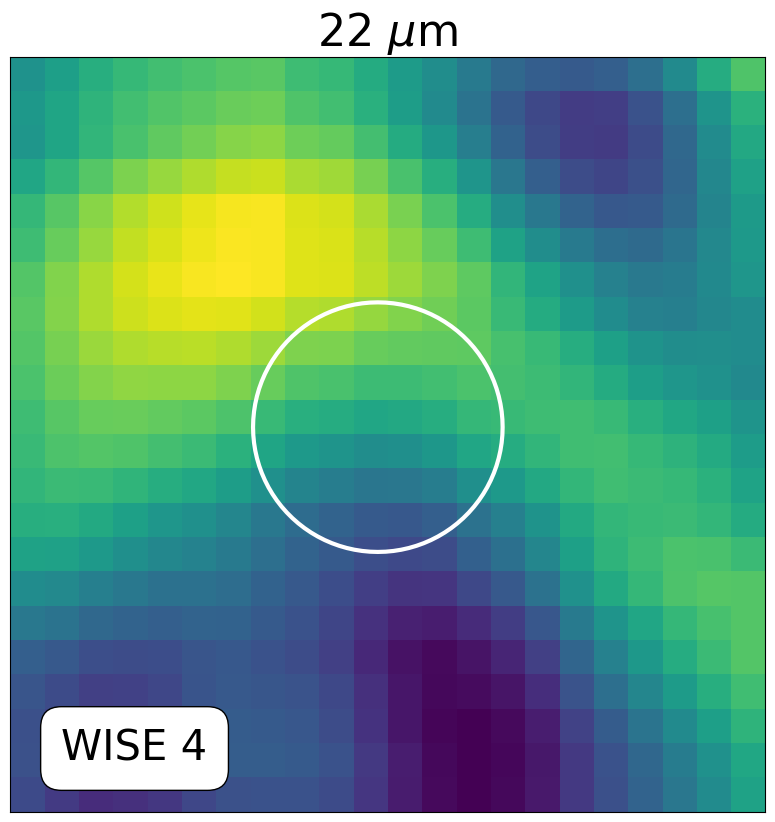} 
    \caption{Cutouts of optical-to-MIR images for J1135 centred on the\textit{ Herschel} position. We plot a white circle of 5 arcsec radius for reference. The postage stamps are 30$\times$30 arcsec}
    \label{fig:multiband_cutouts}
\end{figure*}

The multi-band (optical-to-mid IR) image cut-outs of J1135 are reported in Figure \ref{fig:multiband_cutouts}. A faint emission at $\sim 4$ $\sigma$ emerges starting from the VIKING H-band and is detected in both IRAC channels with a S/N $\gtrsim 6$, but the angular resolution is not sufficient to resolve any lensing-features (e.g arcs) in the NIR/MIR regime. 
Flux densities are estimated by performing aperture photometry with an aperture diameter of 2 arcsec for NIR VIKING images and 6 arcsec for \textit{Spitzer}/IRAC images. 
Table \ref{tab:photometry} summarises the photometry for J1135, we report upper limits for non-detections (i.e. emission with S/N $\lesssim 3$ ) .

\begin{table}
\centering
\caption{Photometric data for J1135. We show references for flux densities (in mJy) taken from the catalogues described in Sec \ref{sec:multiband_obs} while the remaining values are extracted through aperture photometry. Upper limits are reported at the $3\sigma$ level.}
\label{tab:photometry}
\begin{tabular}{ccc} 
\hline
Wavelength & Flux density & Instrument      \\ 
($\mu$m) & (mJy) & \\
\hline
0.47                                                      & $\lesssim$0.09$\times 10^{-3}$                                                 & HSC/g           \\
0.61                                                      & $\lesssim$0.17$\times 10^{-3}$                                                 & HSC/r           \\
0.77                                                      & $\lesssim$0.26$\times 10^{-3}$                                                   & HSC/i           \\
0.89                                                      & $\lesssim$0.41$\times 10^{-3}$                                                    & HSC/z           \\
0.97                                                      & $\lesssim$0.43$\times 10^{-3}$                                                    & HSC/y           \\
1.15                                                        & $\lesssim$0.91$\times 10^{-3}$                                                   & HST/WFC3        \\
1.64                                                       & (6.9$\pm$1.2)$\times 10^{-3}$                                         & VIKING/H        \\
2.15                                                       & (7.8$\pm$1.2)$\times 10^{-3}$                                         & VIKING/Ks       \\
3.55                                                       & (37.1$\pm$6.2)$\times 10^{-3}$                                     & Spitzer/IRAC1   \\
4.49                                                       & (58.2$\pm$7.5)$\times 10^{-3}$                                    & Spitzer/IRAC2   \\
11.6                                                         & $\lesssim$0.45                                         & WISE/W3$^1$        \\
22.1                                                        & $\lesssim$3.83                                                       & WISE/W4$^1$         \\
100                                                         & $\lesssim$136.3                                                     & Herschel/PACS$^2$  \\
160                                                         & 151.5$\pm$50.3                                             & Herschel/PACS$^2$   \\
250                                                         & 278.8$\pm$7.4                                                 & Herschel/SPIRE$^{2,3}$   \\
350                                                         & 282.9$\pm$8.2                                                 & Herschel/SPIRE$^{2,3}$  \\
500                                                         & 204.0$\pm$8.6                                                 & Herschel/SPIRE$^{2,3}$   \\
640                                                         & 163.7$\pm$7.9                                              & ALMBA/B8        \\
850                                                         & 118.8$\pm$8.5                                              & SCUBA-2$^4$         \\
1043                                                        & 29.4$\pm$1.4                                                & ALMA/B7         \\
1300                                                        & 16.2$\pm$0.8                                                & ALMA/B6         \\
3450                                                        & 0.71$\pm$0.04                                             & ALMA/B3         \\
48100                                                       & 0.09$\pm$0.01                                              & EVLA/BC         \\
\hline
\end{tabular}

$^1$ From the WISE All-sky Data Release \cite{Wright2010} \\
$^2$ From the HATLAS Data Release 1 catalogue described in \cite{Valiante2016} \\
$^3$ From the HATLAS Data Release 2 catalogue described in \cite{Maddox2017a} \\
$^4$ From the \textit{Herschel} bright sources (HerBS) sample (\citealt{Bakx2018})
\end{table}

\section{Lens modeling and source reconstruction}\label{sec:lens_model}

In order to reconstruct the intrinsic background source morphology we perform lens modeling analysis with the open source Python 3.6+ code \texttt{PyAutoLens} (\citealt{Nightingale2018,Nightingale2021}), which implements the Regularized Semi-Linear Inversion (SLI) Method described in \cite{Warren2003} together with the adaptive source plane pixelization scheme described in \cite{Nightingale2015} adapted to interferometric data as done in \cite{Dye2018,Dye2022}, \cite{Enia2018}, \cite{Massardi2017}, \cite{Maresca2022} and detailed in Appendix \ref{sec:SLI}.

\subsection{Lens model}

In reconstructing the source's light profile we first need to assume a density profile for the mass of the foreground object. The lens is modelled as a Singular Isothermal Ellipsoid (SIE; \citealt{Kormann1994}), i.e. an elliptical power-law density distribution which goes as $\rho \propto r^{-\alpha}$, with $r$ being the elliptical radius and with a fixed slope value $\alpha=2$.
The profile is described by five parameters: the Einstein radius $\theta_{\rm E}$, the lens centroid positions $x_c$, $y_c$, the first and the second ellipticity components of the elliptical coordinate system ($e_x$, $e_y$).
The latter originate from two quantities: the positional angle ($\phi$), defined counter-clockwise from the positive x-axis, and the factor $f= (1 - q) / (1 + q)$ where $q$ is the ratio between the semi-major and semi-minor axis. The final expressions for the elliptical components are:

\begin{align}
    \begin{split}
    e_y = f \times \sin{(2\phi)}, \\
    e_x = f \times \cos{(2\phi)}.
    \end{split}
\end{align}

\texttt{PyAutoLens} performs lens fitting through the nested sampling algorithm \texttt{Dynesty} (\citealt{Speagle2020}) which samples the parameter space and computes the posterior probability distributions for the parameters of a given lens model.

\begin{figure*}
    \centering
	\includegraphics[width=\textwidth]{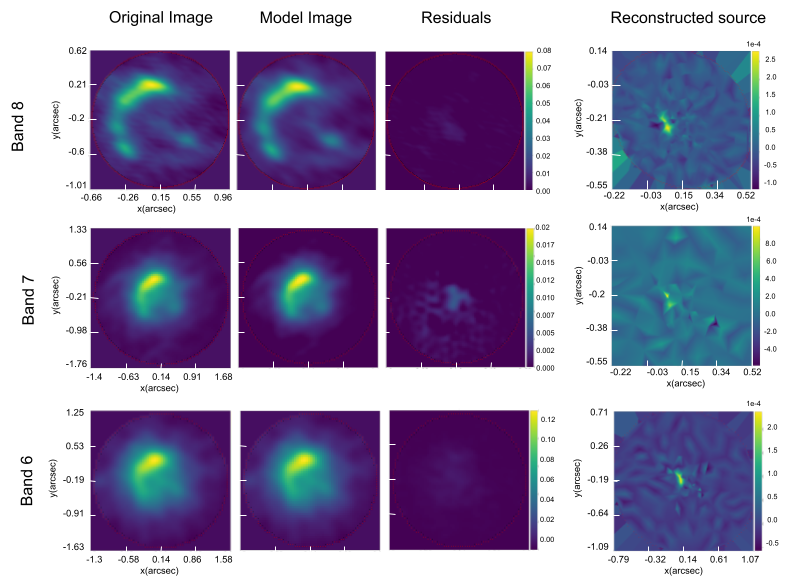}

   \caption{Results of the lens modeling and source reconstruction procedure for continuum data. From the first column to the right: the original ALMA image, the best-fit lensed model image, the fit residuals, and the reconstructed source. The colour bar indicates the surface brightness in units of Jy arcsec$^{-1}$. From the first raw: continuum emission in bands 8, 7, and 6. Note that the y axis is inverted.}
    \label{fig:model_results_continuum}
\end{figure*}

\begin{figure*}
    \centering
	\includegraphics[width=\textwidth]{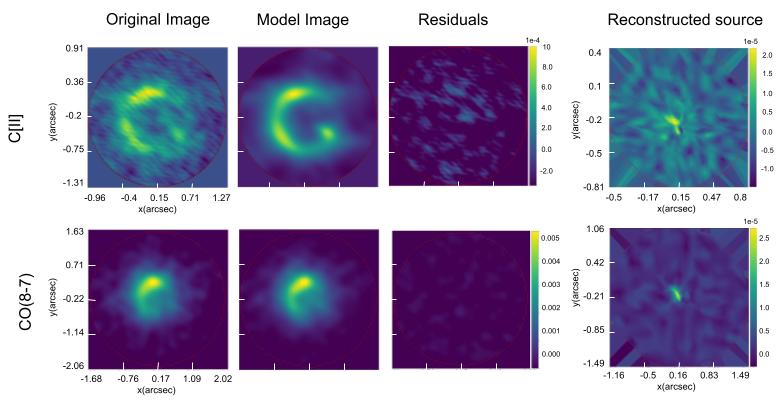}  
   \caption{Same input and output as is Fig. \ref{fig:model_results_continuum} for C[II] and CO(8-7) line data, represented in the first and second raw respectively. }
    \label{fig:model_results_lines}
\end{figure*}

Our searching chain consists in a first non-linear search aimed at setting priors for the lens model, this allows us to exclude regions in the parameter space corresponding to un-physical solutions where the code could get stuck. The best-fit lens model parameters are then used as priors for a second search aimed at initialising the inversion, improving the computational process. A final search is then performed to fully optimise the lens-model parameters. The fit is performed on a number of pixels delimited by a circular mask, where the radius changes according to the resolution of the cleaned ALMA image, in order to obtain a satisfactory fit without exceeding in terms of computational cost.
The output best-fit parameters and their uncertainties are reported in Table \ref{tab:lens_model_best}. Fig. \ref{fig:model_results_continuum} and Fig. \ref{fig:model_results_lines} shows the original lens-plane image, the model image, the residual map and the reconstructed source for the three ALMA continuum bands and the CO(8-7) and C[II] emission lines respectively. Differences in the retrieved physical scales values reflect the heterogeneity of the data adopted in this work, which are the product of different array configurations and angular resolutions.  

\begin{table*}
\begin{center}
\caption{Parameters for the best-fit lens model. $\theta_{\rm E}$ is the Einstein radius measured in arcsec, the lens positions are measured in arcsec and are referred to the centre of the ALMA observation. q and $\phi$ are the axis ratio and the positional lens angle respectively, derived from the elliptical components as described in section \ref{sec:lens_model}. $\phi$ is defined from the positive horizontal axis. }
\label{tab:lens_model_best}
\begin{tabular}{cccccccc}
\hline
 & \multicolumn{1}{c}{} & \multicolumn{1}{c}{$\theta_{\rm E}$} & \multicolumn{1}{c}{$\Delta $x$_L$} & \multicolumn{1}{c}{$\Delta $y$_L$} & \multicolumn{1}{c}{q} & $\phi$                  \\
 &  & (arcsec) & (arcsec) & (arcsec) & & (deg) 
 \\
 \hline
 & Band 8 & $0.460^{+0.001}_{-0.01}$             & $0.163^{+0.001}_{-0.028}$           & $-0.268^{+0.009}_{-0.001}$          & $0.65_{-0.01}^{+0.01}$  & $-33.09^{+0.02}_{-0.03}$  \\
 & Band 7 & $0.471^{+0.002}_{-0.010}$             & $0.165^{+0.008}_{-0.012}$           & $-0.252^{+0.001}_{-0.001}$          & $0.62_{-0.60}^{+0.64}$  & $-33.27^{+0.03}_{-0.04}$  \\
& Band 6 & $0.447^{+0.001}_{-0.007}$             & $0.187^{+0.001}_{-0.021}$           & $-0.213^{+0.008}_{-0.001}$          & $0.57_{-0.01}^{+0.01}$  & $-34.09^{+0.02}_{-0.02}$  \\
& C[II] & $0.468^{+0.058}_{-0.013}$             & $0.159^{+0.003}_{-0.003}$           & $-0.280^{+0.027}_{-0.101}$          & $0.62_{-0.15}^{+0.19}$  & $-17.23^{+0.79}_{-0.70}$  \\
& CO(8-7) & $0.463^{+0.003}_{-0.001}$             & $0.240^{+0.051}_{-0.067}$           & $-0.209^{+0.456}_{-0.002}$          & $0.50_{-0.16}^{+0.12}$  & $-30.87^{+0.17}_{-0.19}$ \\ 
   
\hline
\end{tabular}
\end{center}
\end{table*}

\begin{table*}
\begin{center}
\caption{Properties of the reconstructed source. From the left: magnification factor, de-magnified angular resolution and effective radius for 3$\sigma$ and 5$\sigma$ emission. }
\label{tab:rec_source}
\begin{tabular}{ccccccc}
\hline
 & \multicolumn{1}{c}{} & \multicolumn{1}{c}{$\mu$ } & \multicolumn{1}{c}{$\theta$} &  \multicolumn{1}{c}{R$_{\rm eff, 3\sigma}$} & \multicolumn{1}{c}{R$_{\rm eff, 5\sigma}$}   &   \\
 &  & & (arcsec) & (pc) & (pc) 
 \\
 \hline
& Band 8 &  12.9$\pm$0.2  & 0.02                    & 439.02          & 363.1  \\
& Band 7 & 7.1$\pm$0.1   & 0.08                   & 339.3         & 256.3  \\
& Band 6 &  7.9$\pm$0.1  & 0.09                   & 953.01          &  740.8  \\
& C[II] &  5.9$\pm$0.3    & 0.04                 & 498.7        & 390.4   \\
& CO(8-7) &   8.1$\pm$0.6 &    0.1                & 1378.3          & 1098.2 \\
   
\hline
\end{tabular}
\end{center}
\end{table*}

\begin{figure}
    \centering
    \includegraphics[width=0.45\textwidth]{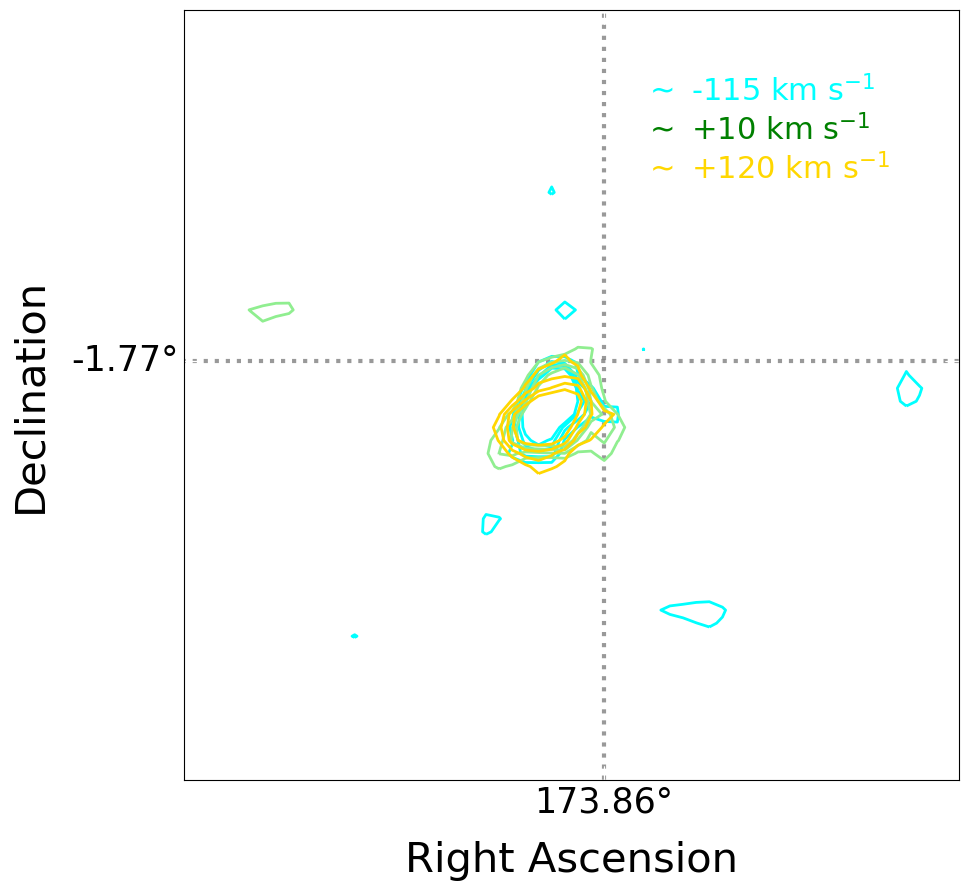}
    \caption{Reconstructed velocity map for CO(8-7) line emission. Contours represent the reconstructed surface brightness for 3 different velocity bins.}
    \label{Fig:velocity}
\end{figure}

Moreover, we reconstruct the velocity map for the CO(8-7) line by dividing and modeling the emission in three different velocity bins. As there is no significant difference in the reconstructed emission in the bins, we cannot claim any indication of rotation or outflow (see Fig. \ref{Fig:velocity}). 

The resulting source physical properties are reported in Table \ref{tab:rec_source}.
The magnification factor is computed as the ratio between the lensed and unlensed surface brightness. The effective radius is computed from the area enclosing all the pixels with signal-to-noise ratio $\gtrsim 3$ and $\gtrsim 5$ in the reconstructed source plane (SP) as $A^{SP} = \pi r^2_{\rm eff}$. The resulting reconstructed source contains only pixels excluded from the masked lensed image area. 
This key information allows us to retrieve the intrinsic properties of the lensed background object. 

\subsection{The lens}\label{sec:lens}

One peculiar aspect of the J1135 gravitational lensed system is the faintness of the foreground object. Indeed, no redshift estimate is available for the lens galaxy and no clear detection is measured from the photometric images, likely due to an insufficient sensitivity and/or angular resolution. As showed in analogous studies and as revealed by HST/NIR high resolution images (e.g \citealt{Negrello2014}), the foreground object usually dominates the emission in those bands, with a progressively higher contribution coming from the background source at higher wavelengths. For this reason, in order to achieve reliable results from the SED-fitting procedure, it is essential to fit and subtract the light profile of the foreground galaxy. 
In this case, however, only a marginal emission ($\lesssim3\sigma$) comes from HST WFC3/F110 data and it is not possible to establish a priori whether it is originated by the lens or by the lensed object.

\begin{figure}
    \centering
    \includegraphics[width=0.5\textwidth]{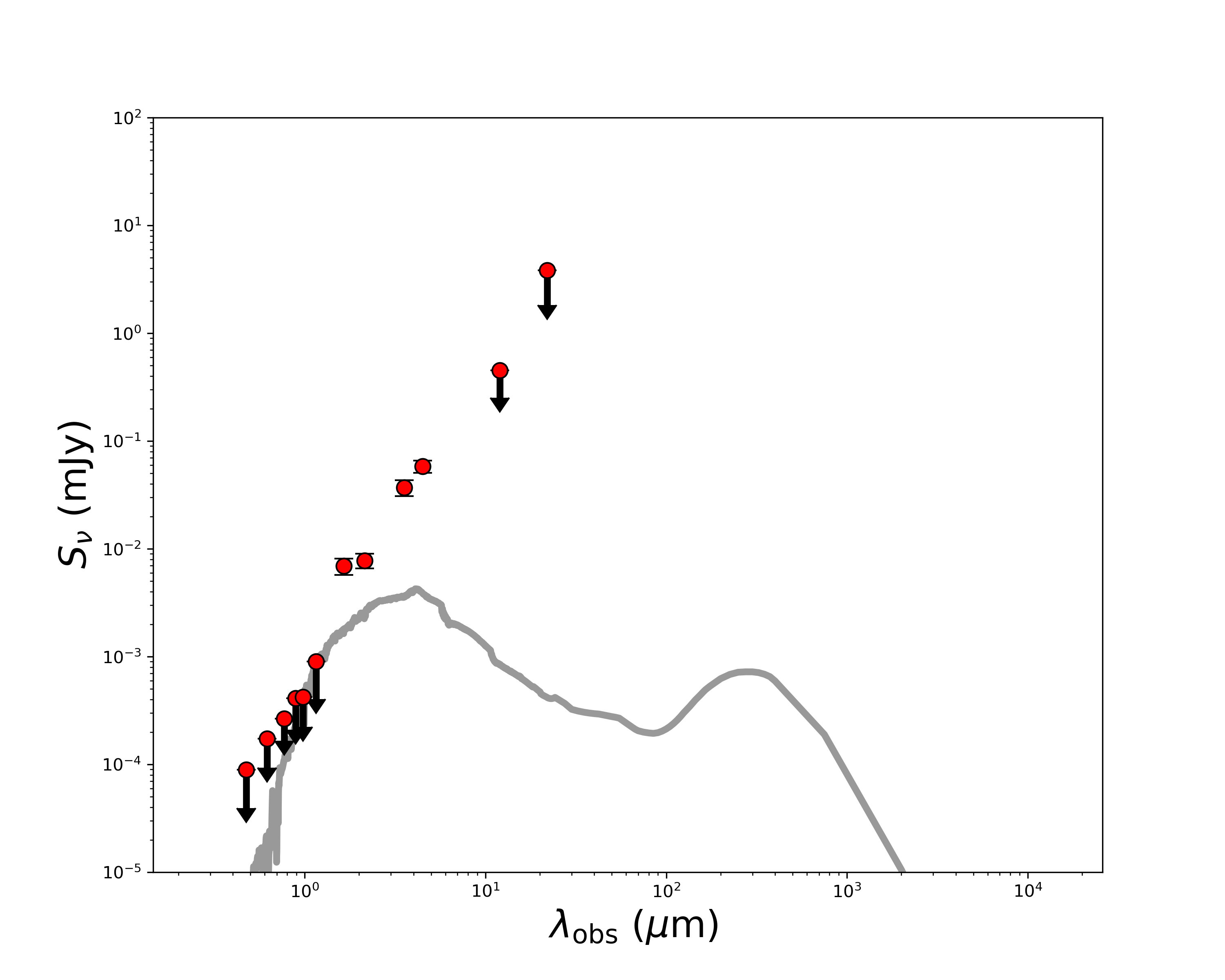}
    \caption{SED template of an passive elliptical galaxy at redshift $\sim1.5$ compared with the photometry of J1135. The flux densities reported in Table \ref{tab:photometry} from HCS/g to SCUBA/850 $\mu$m are represented as red points. Upper limits at 3$\sigma$ are showed as arrows.}
    \label{fig:sed_test}
\end{figure}

We therefore assume the lens to be a massive elliptical, and attribute its faintness to its relatively high redshift (e.g. $z\gtrsim 1.5$). We model the SED of the foreground object according to this assumption and constrain its luminosity by means of the Einstein (total) mass resulting from the lens modeling (M$_E \sim 1.3 \times 10^{11}$ M$_{\odot}$). Specifically, we adopt the template for an elliptical galaxy with $2$ Gyr age from the SWIRE library (\citealt{Polletta2007}). The resulting SED of this template overlapped with the photometry reported in Table \ref{tab:photometry} is showed in Fig. \ref{fig:sed_test}.
We find the contribution from the lens to be negligible for the flux densities from the H and Ks VIKING bands up to the higher wavelengths, hence no lens-subtraction is needed. 
The situation is less clear for the marginal HST WFC3/F110 detection and for this reason, we consider this value as an upper limit.

\section{Far Infrared SED and radio properties}\label{sec:fir_sed}

\begin{figure}
    \centering
    \includegraphics[width=0.5\textwidth]{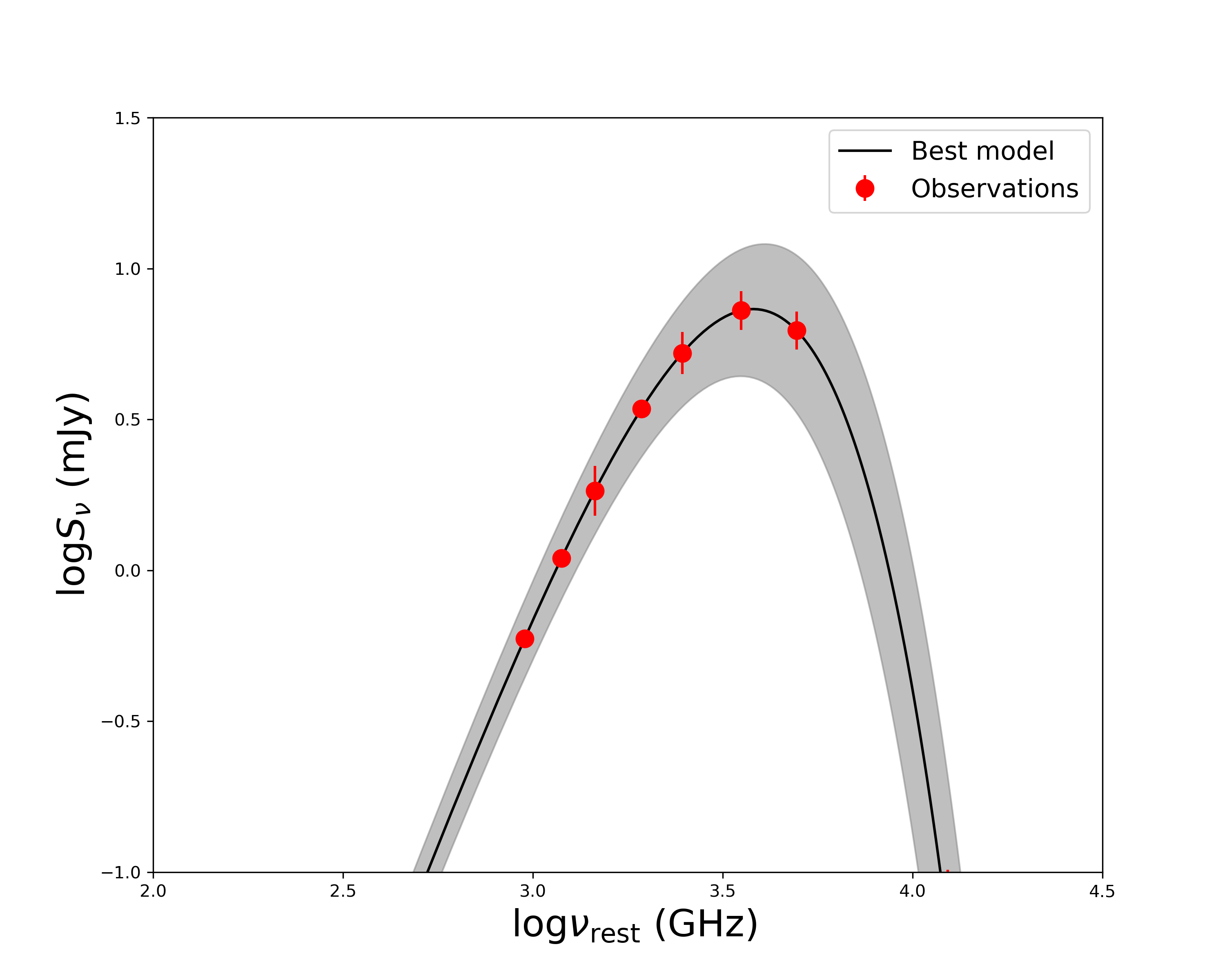}
    \caption{Best-fit FIR to sub-mm rest-frame SED of J1135. Red points are the observed flux densities and errors and the black line is the best-fitting modified black body spectrum. The grey shaded area represents the 68$\%$ confidence interval for the best-fit model. }
    \label{fig:fir_sed}
\end{figure}

We compute the intrinsic Far Infrared Luminosity (FIR), defined in the wavelength range of 8-1000 $\mu$m, by de-magnifying and fitting the FIR-to-sub-mm flux densities available for G12H43. We consider the \textit{Herschel}/PACS and \textit{Herschel}/SPIRE photometry from the \cite{Negrello2017} sample, we include the SCUBA-2 880$\mu$m integrated flux density reported in \cite{Bakx2018}, and the flux density value measured in the 0.64, 1.04, and 1.3 mm continuum ALMA image.
We use a single temperature modified black body under the optically-thin approximation, where the dust emissivity index is fixed at $\beta =$ 1.5 (\citealt{Nayyeri2016}; \citealt{Negrello2017}), while the spectrum normalisation and the dust temperature (T$_{\rm dust}$) are kept as free parameters. The model ($S_{\nu, best}$) which minimises the $\chi^2$ is then integrated over the wavelength range 8-1000 $\mu$m as follows: 
\begin{equation}
    L_{\rm FIR}= \frac{4\pi D_L^2}{(1+z)} \int^{1000\mu \rm m}_{8 \mu \rm m} S_{\nu, best} d\nu.
\end{equation}
The best-fit spectral energy distribution is represented in Figure \ref{fig:fir_sed}, corresponding to a dust temperature T$_{\rm dust} = 41.1 \pm 2.9$ K and to a resulting far infrared luminosity of $L_{\rm FIR} = \log(L/ L_{\odot}) = 12.91 \pm 0.01$.

By assuming a power law spectrum $S_{\nu}=\nu ^{\alpha}$ with average radio spectral index $\alpha=-0.7\pm 0.14$, we compute the rest-frame radio luminosity $L_{\rm 1.4 GHz}$ at 1.4 GHz as: 

\begin{equation}\label{eq:L_radio}
    L_{\nu,e}= \frac{4 \pi D_{L}^2(z)}{(1+z)^{1+\alpha}} \left( \frac{\nu_e}{\nu_o} \right)^{\alpha} S_{\nu,o},
\end{equation}

where $S_{\nu,o} \propto \nu ^{\alpha}$ is observed monochromatic flux density at 6 GHz corrected for a magnification factor computed as the mean of the magnification factors in output from the lens modeling of the ALMA continuum emission. $\nu_e$ and $\nu_o$ are the emitted and the observed frequency and $D_L$ is the luminosity distance. We obtain $\log L_{\rm 1.4 GHz} = 24.2\pm 0.2$ W $\hbox{Hz}^{-1}$.
Finally, we derive the far-infrared/radio correlation as: 

\begin{equation}\label{eq:q_ir}
q_{\rm FIR}=\log\left(\frac{L_{\rm FIR}[\hbox{W}]/3.75\times 10^{12}}{L_{1.4\,\rm GHz} [\hbox{W}\,\hbox{Hz}^{-1}]}\right),
\end{equation}

inferring a value of $q_{\rm FIR} = 2.69\pm0.41$.

\section{SED fitting}\label{sec:SED fitting}

By correcting the available photometric information for the magnification factor we can retrieve the \textit{intrinsic} physical properties of J1135. 
To achieve this goal, we perform Spectral Energy Distribution (SED) fitting with the e Code Investigating GAlaxy Emission (CIGALE, \citealt{Boquien2019}). CIGALE is a Python SED fitting code able to reproduce broad-band uv-to-radio photometric data according to the energy balance (i.e the energy coming from the stellar uv-NIR emission is the same as the one re-emitted by the dust in the MIR and FIR regime).
The main physical properties are estimated by comparing the observed galaxy SED with the modelled one by means of a $\chi^2$ and bayesian statistics. We exploit the available broad-band photometry described in Sect. \ref{sec:multiband_obs} and the continuum ALMA emission, including a 3$\sigma$ upper limit for non-detections. For low-resolution data, we correct the flux density values for the average magnification described in the previous sectiion. As described in Sect. \ref{sec:lens}, we adopt the assumption that the observed photometry belongs only to the lensed source. In the following, we describe the modules adopted for the SED-fitting procedure.

The stellar emission is computed following the \cite{Bruzual2003} (BC03) population synthesis models, associated with a \cite{Chabrier2003} Initial Mass Function (IMF) and metallicity values of Z= 0.004, 0.008, 0.02, 0.05. 
We assume a delayed star formation history, which predicts a nearly linear increase of the SFR:

\begin{equation}
    SFR(t) \propto \frac{t}{\tau^2} \times \exp{\left(-\frac{t}{\tau}\right)} \hspace{1cm} \text{for} \hspace{0.5cm}  0\leq t \leq t_0, 
\end{equation}

where $t_0$ being the age of the onset of star formation, and $\tau$ the time at which the SFR peaks. 

In order to model the effect of the dust attenuation on FUV-optical light we adopt the modified \cite{Charlot2000} prescriptions, where the attenuation is age-dependent and described by two different power-laws, one for the ISM and one for the Birth Clouds (BC). The attenuation slopes are assumed to be -0.7 and the V-band attenuation is computed as:

\begin{equation}
    \mu = \frac{A_{\rm V}^{\rm ISM}}{A_{\rm V}^{\rm ISM} + A_{\rm V}^{\rm BC}}.
\end{equation}

In our analysis we assume $A_{\rm V}^{\rm ISM}$ spanning from 0.3 to 5.0 and a $\mu$ spanning from 0.3 to 0.6. 

Following \cite{Draine2007}, dust emission is modelled as two separated components: a diffuse one, illuminated with a single radiation field ($U_{min}$) originated by a general stellar population; and a second component is closely associated to regions in which the star-formation occurs, heated by a variable radiation field described with a power-law profile with index $\alpha$ and defined between two values $U_{min}$ and $U_{max}$. In particular, we use the most recent and refined version of this model which accounts also for dust-mass renormalisation (\citealt{Draine2014}).

The resulting value of the FIRRC parameter q$_{\rm FIR}\sim 2.7$ computed in Sect. \ref{sec:fir_sed}, is used as a prior for the CIGALE synchrotron module to fit radio flux density at 6 GHz assuming a fixed slope ${\alpha}=$-0.7 as in Eq. \ref{eq:L_radio}. 

\begin{figure*}
    \centering
    \includegraphics[width=0.65\textwidth]{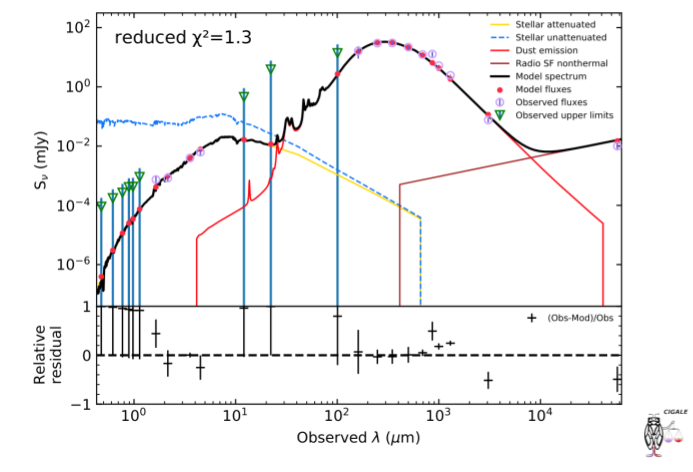}
    \caption{Best-fit UV to radio observed-frame SED of J1135. Green arrows are 3$\sigma$ upper limits, purple circles are the observed flux densities and errors. The black line is the best-fitting modified black body spectrum.}
    \label{fig:sed}
\end{figure*}

The best-fit model is presented in Fig. \ref{fig:sed} and the resulting best physical properties are summarised in Table \ref{tab:SED_fitting_res}.

\begin{table}
\centering
\caption{Best-fit output parameters from CIGALE. From the first raw: dust luminosity, dust mass, star formation rate and stellar mass.}
\label{tab:SED_fitting_res}
\begin{tabular}{cl} 
\hline

\multicolumn{2}{c}{SED-fitting results }                                                               \\ 
\hline
\begin{tabular}[c]{@{}c@{}}$\log{\rm L}_{\rm dust} $ (L$_{\odot}$)\end{tabular}  & 13.01 $\pm$ 0.06    \\
\begin{tabular}[c]{@{}c@{}}$\log{\rm M}_{\rm dust}$ (M$_{\odot}$)\end{tabular}    & 9.21 $\pm$ 0.05     \\
\begin{tabular}[c]{@{}c@{}} $\log{\rm SFR}$ (M$_{\odot}$ yr$^{-1}$)\end{tabular}              & 2.92 $\pm$ 0.08  \\
\begin{tabular}[c]{@{}c@{}}$\log{\rm M}_{\rm \star}$ (M$_{\odot}$)\end{tabular} & $\lesssim$ 11.73      \\
\hline
\end{tabular}
\end{table}

\section{Discussion}\label{sec:discussion}

Taking advantage from the SED-fitting results, we are able to investigate the ISM conditions of J1135 and its evolutionary state.

\subsection{Stellar and gas masses}\label{sec:masses}
The bunch of available data allow us to estimate the gas content by adopting several empirical calibrators. First, we directly estimate the gas mass from the C[II] following \cite{Zanella2018}, we assume $\alpha_{\rm C[II]} \equiv M_{\rm gas} / L_{\rm C[II]} = 22M_{\odot} /L_{\odot}$, which is calibrated on starburst galaxies spanning a redshift range $z\sim 2-6$.
Secondly, in order to estimate the molecular gas content (M$_{\rm H_2}$) we derive the $L'_{\rm CO(1-0)}$ from the de-magnified $L'_{\rm CO(8-7)}$ luminosity. We then follow \cite{Fujimoto2022} adopting a conversion factor of  $L'_{\rm CO(1-0)}=1.5 L'_{\rm CO(7-6)}$ estimated for high redshift starburst galaxies in literature (e.g. \citealt{Riechers2013}). This conversion factor is referred to a different transition, corresponding to higher luminosity values of the CO-SLED (\citealt{Yang2017}), for this reason the resulting value of $L'_{\rm CO(1-0)}=1.2\times 10^{10}$ K km $\hbox{s}^{-1}$pc$^2$ is considered as an upper limit. This estimate is consistent with the value of $L'_{\rm CO(1-0)} \sim 1.5 \times 10^{10}$ found by \cite{Harris2012} adopting an indicative magnification factor of 10. We then compute the molecular gas mass assuming two different values of $\alpha_{\rm CO} = 0.8- 4.6$. The value of $\alpha=0.8$ is calibrated from local ULIRGs with super-solar metallicity (\citealt{Downes1998}), while the higher value is calibrated in the Milky Way (\citealt{Solomon1991}).

The molecular gas ISM content can also be estimated by means of the empirical calibration (\citealt{Scoville2017}) as $\alpha \equiv < L_{\nu850\mu m} / M_{gas} > = 6.7 \pm 1.7 \times 10^{19}$ erg s$^{-1}$ Hz$^{-1}$ M$_{\odot}^{-1}$.

Finally, we convert the dust content resulting from the SED fitting into gas assuming a variable gas-to-dust ratio of $\delta_{\rm GDR}=30-92$ referred to typical solar and super solar metallicity following \cite{Magdis2012} and \cite{Fujimoto2022}.
The values obtained for the molecular gas masses are reported in Table \ref{tab:mol_mass}.

\begin{table}
\centering
\caption{Values for the molecular mass computed from different calibrators.}
\label{tab:mol_mass}
\begin{tabular}{lc} 
\hline
Calibrator         & $\log$ M$_{\rm gas}$ (M$_{\odot}$)  \\ 
\hline
C[II]              & 11.04 $\pm $0.3                     \\
CO(1-0)            & $\lesssim$ (10-10.8)                \\
850 $\mu$m         & 11.5$ \pm$0.2                       \\
$\alpha_{\rm GDR}$ & (10.51-11.04)$\pm $0.05             \\
\hline
\end{tabular}
\end{table}

The stellar mass estimate in output from the SED fitting must be considered as an upper limit. Indeed given the lack of a clear detection in NIR images it is not possible to correctly estimate the contribution coming from the lens (see Section \ref{sec:lens} for a further discussion). Moreover, the dark-nature of this object hinders a complete sampling of the optical and NIR part of the SED. Aside from the value reported in Table \ref{tab:SED_fitting_res}, we compute the stellar mass assuming a typical stellar-to-dust mass ratio of $\delta_{\rm SDR}\approx 100$, obtaining a value of $M_*^{\rm STD} \sim 2\times 10^{11}$ M$_{\odot}$, in agreement with the SED fitting estimate.

\subsection{ISM properties}

From the gas mass values reported in Sect. \ref{sec:masses}, we estimate a depletion timescale of $\tau_{\rm depl} \sim 10^8$ yr. Moreover, the inferred stellar mass implies $\tau_{\rm SFR} \sim 10^8$ yr, indicative of a young galaxy, offset from the main sequence locus of star-forming galaxies at z$\sim$3 (\citealt{Speagle2014}). Our results are consistent with the expectations reported in \cite{Vishwas2018}, where the analysis of the Lyman continuum photons required to sustain the luminosity of the O[III] 88 $\mu$m line pointed out to the presence of young and massive stars ionising the surrounding HII regions. 
The same authors found no significant AGN contribution from the SED analysis, consistent with what we infer from the FIRRC (q$_{\rm FIR}\approx 2.7$), which is indicative of a star-formation dominated object.

The hypothesis of J1135 being a compact starburst is also supported by the source reconstruction of the highest angular resolution ALMA continuum emission at 640 $\mu$m shown in Figure \ref{fig:model_results_continuum}, where the effective radius reaches values of $\sim 400$ pc. 
Similar physical scales are reached by the C[II] line emission (see Table \ref{tab:rec_source} and Fig. \ref{fig:model_results_lines}). The C[II] is a fine structure line predominantly originated from high$-z$ photon-dominated regions (PDR) and is typically used as a cool interstellar gas tracer and as a SFR estimator (see \citealt{Casey2014} for a review). A well known deficit in the C[II]/FIR ratio is observed in both nearby (e.g. \citealt{Luhman2003}, \citealt{DiazSantos2017}, \citealt{Smith2017}) and high-redshift star-forming galaxies (\citealt{Stacey2010}, \citealt{Gullberg2015}). This drop is found to reach very low values ($L_{\rm C[II]}/L_{\rm FIR} \approx 10^{-4} $) in spatially resolved studies (e.g. \citealt{Lagache2018}, \citealt{Gullberg2015}, \citealt{Rybak2019}).
For J1135, we infer a C[II]/FIR ratio of $L_{\rm C[II]}/L_{\rm FIR} \approx  5.4 \times 10^{-4}$. Similar values are found for other strongly lensed galaxies among the HATLAS sample. For example, \cite{Rybak2020} reported a deficit down to $\sim 3 \times 10^{-4}$ for spatially resolved ALMA data of SDP.81 (\citealt{Partnership2015}, \citealt{Rybak2015a,Rybak2015}, \citealt{Dye2015}, \citealt{Swinbank2015}, \citealt{Tamura2015a}, \citealt{Hatsukade2015a}, \citealt{Hezaveh2016}) at $z=3.042$. \cite{Lamarche2018} found similar values ($\sim$ 2 $\times$ 10$^{-4}$) for SDP.11 at $z=1.7$, even though our galaxy shows a more compact morphology in the C[II] emission with respect to other objects.
From R$_{\rm eff, 640 \mu m}$ we infer a star-formation surface density of $\rho_{\rm SFR} \sim 1600$ M$_{\odot}$ yr$^{-1}$ kpc$^{-2}$, which is on the verge of the Eddington limit for a radiation pressure supported starburst (\citealt{Andrews2011}, \citealt{Simpson2015}). This value, is compatible with the possible explanation of the deficit to be attributed to a lower increase of the C[II] emission with respect to the FIR.

\subsection{Evolutionary interpretation}

By inspecting the HST/WFC3 image we find no evidence for galaxy companions of J1135 within a radius of at least $\sim 5$ arcsec, corresponding to $\sim 40$ kpc, so that we can exclude a merger-induced origin of the starburst.
Thus the ISM conditions and the physical properties discussed so far can be interpreted in the light of in-situ galaxy formation scenarios (\citealt{Lapi2014,Lapi2018}, \citealt{Mancuso2017}, \citealt{Pantoni2019}). 
In particular, the properties of J1135 are consistent with a \textit{compaction} phase (see Fig. 3 in \citealt{Lapi2018}) in which the dust-enshrouded star-formation activity increases at an almost constant rate in the inner regions of the galaxy where the stellar mass is being accumulated. At this stage, the in-situ scenario envisages the galaxy to be an off-main sequence object in a early evolutionary stage, which will eventually move towards the main-sequence locus as the stellar mass content increases. Finally, the star formation will either progressively decrease as the galaxy exhaust its gas reservoir or will be abruptly stopped by the action of the feedback from an AGN (\citealt{Mancuso2017}).

\section{Summary and conclusions}\label{sec:conclusions}

In this work we have investigated the nature of the strongly-lensed galaxy HATLASJ113526.2-01460 (namely, J1135) at redshift $z\approx 3.1$, discovered by the \textit{Herschel} satellite in the GAMA 12$^{\rm th}$ field of the \textit{Herschel}-ATLAS survey. We have performed detailed lens modeling and have reconstructed the source morphology in three different (sub-)mm continuum bands, and in the spectral emission of the C[II] and CO(8-7) lines. We have also exploited a wealth of photometric ancillary data to perform broadband SED-fitting and to retrieve intrinsic (i.e., corrected for magnification) physical properties. Our main findings are summarized below:

\begin{itemize}

\item The lens modeling indicates that the foreground lens is constituted by a (likely elliptical) galaxy with mass $\gtrsim 10^{11}\, M_\odot$ at $z\gtrsim 1.5$, while the source is found to be an optical/NIR dark, dusty star-forming galaxy whose (sub-)mm continuum and line emissions are amplified by factors $\mu\sim 6-13$.

\item The emission of J1135 is extremely compact, with sizes $\lesssim 0.5$ kpc for the star-forming region and $\lesssim 1$ kpc for the gas component, with no clear evidence of rotation or of ongoing merging events.

\item J1135 features a very high star-formation rate $\gtrsim 10^3\, M_\odot$ yr$^{-1}$, that given the compact sizes is on the verge of the Eddington limit for starbursts. The radio luminosity at $6$ cm from available EVLA observations is consistent with the star-formation activity, so that no significant contribution from a central AGN is emerging (see also \citealt{Vishwas2018}).

\item J1135 is found to be extremely rich in gas $\sim 10^{11}\, M_\odot$ and dust $\gtrsim 10^9\, M_\odot$. The stellar content $\lesssim 10^{11}\, M_\odot$ places J1135 well above the main sequence of starforming galaxies, indicating that the starburst is rather young with an estimated age $\sim 10^8$ yr, and that the stellar mass should at least double before star formation is quenched. 

\item The properties of J1135 can be consistently explained in terms of in-situ galaxy formation and evolution scenarios as typical of a rather young dusty starforming galaxy caught in the compaction phase.

\end{itemize}

In the next future, observations coming from the James Webb Space Telescope (JWST) will be crucial to shed further light on the nature of this obscured object and its foreground lens in the near- and mid-IR regime.
Moreover, X-Ray follow-up, coupled with the available ALMA data, are required to establish the presence of the dust-enshrouded AGN and to better investigate the interplay between star-formation and the nuclear activity (\citealt{Massardi2017}). 

\section{Acknowledgments}

This paper makes use of the following ALMA data: 2016.1.01371, 2017.1.01694.S, 2018.1.00861.S. ALMA is a partnership of ESO (representing its member states), NSF (USA) and NINS (Japan), together with NRC (Canada), MOST and ASIAA (Taiwan), and KASI (Republic of Korea), in cooperation with the Republic of Chile. The Joint ALMA Observatory is operated by ESO, AUI/NRAO and NAOJ. We acknowledge financial support from the grant PRIN MIUR
2017 prot. 20173ML3WW 001 and 002 ‘Opening the ALMA window on the cosmic evolution of gas, stars, and supermassive black holes’. AL is supported by the EU H2020-MSCAITN-2019 project 860744 ‘BiD4BEST: Big Data applications for Black hole Evolution STudies’.

\appendix

\section{Adapting the SLI method to interferometric visibilities}\label{sec:SLI}

The SLI formalism can be extended also to interferometry (\citealt{Dye2018}, \citealt{Enia2018}, \citealt{Maresca2022}), modeling a set of visibility data, i.e. the result of the correlation of signals coming from an astrophysical source and collected by the antennae array, whose Fourier transform gives the source surface brightness distribution.
Performing an inversion directly on the Fourier space (or \textit{uv}-plane) circumvents the issue of dealing with artifacts and noise correlation arising in the image as a consequence of a poor sampling of the \textit{uv}-plane.

Following a similar formalism with respect to the one used in \cite{Dye2018}, we introduce the rectangular matrix f$_{ij}$ containing the fluxes of the \textit{i-th} pixel in the source plane and the respective \textit{j-th} image-plane pixel.
Analogously, complex visibilities from the lensed image are collected rectangular matrix g$_{ij}$, which are the Fourier transform of the \textit{i} source pixels in unit surface brightness computed at the \textit{j-th} visibility point in the \textit{uv}-plane. For each \textit{j-th} visibility corresponding to the source pixel surface brightnesses $s_i$, the model visibility set can be described as $\sum_i s_i g_{ij}$. 

Given a set of observed visibilities $V_{obs}$, the merit function can be described as:

\begin{equation}
    G= \frac{1}{2} \chi^2 =  \frac{1}{2} \sum_{j=1}^{J}\left( \frac{\left| \sum_{i=1}^{I} s_i g_{ij}-V_{obs,j}\right|^2}{\sigma_j^2}\right) + \lambda \frac{1}{2} \textbf{S}^T\textbf{HS}, 
\end{equation}

computed over a total of I Delaunay pixels and J visibilities. $\sigma_j$ are the 1$\sigma$ uncertainties on the observed visibilities rescaled adopting the CASA task \texttt{statwt} to match their absolute value. The last term in the expression describes the \textit{regularization}, where $\lambda$ is a constant determining the strength of the regularization, and \textbf{H} the regularization matrix.
The values $s_i$, represented by the vector \textbf{S} which best reproduces the observed image-plane visibilities, can therefore be derived minimizing the merit function $G$. The solution to this linear problem is given by:

\begin{equation}
    \textbf{S} = [\textbf{F} + \lambda \textbf{H}]^{-1} \textbf{D},
\end{equation}
where \textbf{F} and \textbf{D} are respectively the matrices $F_{ij} = \sum^J_{n=1} (g_{in}^{\mathbb{R}}g_{jn}^{\mathbb{R}}+ g_{in}^{\mathbb{I}}g_{jn}^{\mathbb{I}}/\sigma^2_n$) and $D_{i}= \sum^J_{n=1} (g_{in}^{\mathbb{R}}V_{n}^{\mathbb{R}}+ g_{in}^{\mathbb{I}}V_{n}^{\mathbb{I}}/\sigma_n^2)$.

For a fixed mass model, the image plane pixels are traced back to the source plane and grouped together by means of a k-clustering algorithm, comparing each source-pixel with the neighbors sharing a direct vertex. This procedure results in new source plane's centres, used to trace a Delaunay grid.
When dealing with a large number of visibilities, the computational efficiency and the memory costs are greatly improved by performing non-uniform Fast Fourier Transform (NUFFT) algorithm, implemented in \texttt{PyAutoLens} exploiting the \texttt{PyNUFFT} (\citealt{Lin2018}) library and the linear algebra package \texttt{PyLops} (\citealt{Ravasi2020}).

\bibliography{Giulietti22}{}
\bibliographystyle{aasjournal}

\end{document}